\documentclass[lettersize,journal]{IEEEtran}

\IEEEoverridecommandlockouts
\usepackage[a4paper, left=13mm, right=13mm, bottom=43mm, top=19mm]{geometry}

\usepackage[caption=false,font=normalsize,labelfont=sf,textfont=sf]{subfig}
\usepackage{textcomp}
\usepackage{stfloats}
\usepackage{verbatim}
\usepackage{cite}
\usepackage{amsmath,amssymb,amsfonts}
\usepackage{algorithmic}
\usepackage{graphicx}
\usepackage{float}
\usepackage{textcomp}
\usepackage{xcolor}
\usepackage{bbm}
\usepackage{gensymb}
\usepackage{comment}
\usepackage[acronym]{glossaries}

\usepackage{adjustbox}
\usepackage{tikz}
\usetikzlibrary{shapes,arrows}
\usetikzlibrary{positioning}
\usetikzlibrary{decorations.text}

\newcommand{\ones}[1]{\mathbbm{1}\left(#1\right)}

\renewcommand{\Pr}[2]{\mathbb{P}_{#1}\left(#2\right)}
\newcommand{\esp}[2]{\mathbb{E}_{#1}\left[#2\right]}
\newcommand{\restesp}[2]{\mathcal{E}^{#1}\left[#2\right]}

\renewcommand{\exp}[1]{\text{exp}\left(#1\right)}

\renewcommand{\Im}[1]{\text{Im}\left\{#1\right\}}

\newcommand{\PDF}[2]{f_{#1}\left(#2\right)}

\newcommand{\CCDF}[2]{\bar{F}_{#1}\left(#2\right)}

\newcounter{CorrCounter}
\newcounter{LemmaCounter}
\newcounter{PropCounter}

\newcommand{\corrolary}[3]{
\refstepcounter{CorrCounter}
\vspace*{0.2cm}
\noindent\textit{\textbf{Corrolary \theCorrCounter:} #2}\smallskip

\noindent\textit{Proof:} #3\\\null\hfill$\blacksquare$
\vspace*{0.2cm}
\label{corr:#1}
}

\newcommand{\lemma}[3]{
\refstepcounter{LemmaCounter}
\vspace*{0.2cm}
\noindent\textit{\textbf{Lemma \theLemmaCounter:} #2}\smallskip

\noindent\textit{Proof:} #3\\\null\hfill$\blacksquare$
\vspace*{0.2cm}
\label{lemma:#1}
}

\newcommand{\proposition}[3]{
\refstepcounter{PropCounter}
\vspace*{0.2cm}
\noindent\textit{\textbf{Proposition \thePropCounter:} 
#2}\smallskip

\noindent\textit{Proof:}
#3\\\null\hfill$\blacksquare$
\vspace*{0.2cm}
\label{prop:#1}
}

\newcommand{\RCM}[1]{\Psi_{#1}}

\newcommand{\LT}{\mathcal{L}}

\newcommand{\Pv}{{P_V}}
\newcommand{\Pvm}{P_{V\text{m}}}
\newcommand{\PvM}{P_{V\text{M}}}
\newcommand{\Pl}{{P_L}}
\newcommand{\Plm}{P_{L\text{m}}}
\newcommand{\PlM}{P_{L\text{M}}}

\newcommand{\psiVT}{\psi_\text{VT}}
\newcommand{\psiVR}{\psi_\text{VR}}
\newcommand{\psiLT}{\psi_\text{LT}}

\newcommand{\Gr}{G_R}
\newcommand{\Gc}{G_C}
\newcommand{\Gi}{G_I}

\newcommand{\rhoR}{\rho_R}
\newcommand{\rhoC}{\rho_C}
\newcommand{\rhoI}{\rho_I}

\newcommand{\PhiR}{\boldsymbol{\Phi}_R}
\newcommand{\PhiI}{\boldsymbol{\Phi}_I}
\newcommand{\PhiC}{\boldsymbol{\Phi}_L}

\newcommand{\lambdaR}{\lambda_R}
\newcommand{\lambdaI}{\lambda_I}
\newcommand{\lambdaC}{\lambda_L}
\newcommand{\lambdaV}{\lambda_V}
\newcommand{\lambdaL}{\lambda_L}

\newcommand{\rR}{r_R}
\newcommand{\rC}{r_C}
\newcommand{\rRmin}{r_{R\text{min}}}
\newcommand{\rRmax}{r_{R\text{max}}}
\newcommand{\rImin}{r_{I\text{min}}}
\newcommand{\rCmin}{r_{C\text{min}}}

\newcommand{\Srmin}{S_{R\text{min}}}
\newcommand{\Srmax}{S_{R\text{max}}}

\newcommand{\Scmax}{S_{C\text{max}}}

\newcommand{\dI}{d_I}
\newcommand{\dC}{d_C}

\newcommand{\dB}{\text{dB}}

\newcommand{\GHz}{\text{GHz}}

\newcommand{\PLF}{L}
\newcommand{\PLE}{\alpha}
\newcommand{\FI}{\beta}

\newcommand{\fc}{f_c}

\newcommand{\SrRCS}{S_{R,\text{RCS}}}

\newcommand{\Sr}{S_R}
\newcommand{\Sc}{S_C}
\newcommand{\I}{I}

\newcommand{\RCS}{\sigma}

\newcommand{\RPG}{\kappa}

\newcommand{\threshRd}{\gamma}
\newcommand{\threshRs}{\theta'}
\newcommand{\threshC}{\theta}

\renewcommand{\d}{\text{d}}

\newcommand{\PFAcut}{P_{\text{FA}}}
\newcommand{\PDcut}{P_\text{D}}
\newcommand{\PScut}{P_\text{S}}
\newcommand{\PFA}{\mathcal{P}_{\text{FA}}}
\newcommand{\PD}{\mathcal{P}_\text{D}}
\newcommand{\PS}{\mathcal{P}_\text{S}}
\newcommand{\PC}{\mathcal{P}_\text{C}}
\newcommand{\rateCCDF}{\mathcal{R}}
\newcommand{\rate}{\bar{\mathcal{R}}}
\newcommand{\JRDCCP}{\mathcal{D}}
\newcommand{\JRDCCPmax}{\mathcal{D}_\text{M}}
\newcommand{\JRDCCPmin}{\mathcal{D}_\text{m}}
\newcommand{\JRSCCPmax}{\mathcal{S}_\text{M}}
\newcommand{\JRSCCPmin}{\mathcal{S}_\text{m}}
\newcommand{\JRSCCP}{\mathcal{S}}
\newcommand{\PFAPIC}{\mathcal{P}_{\text{FA}}^\text{IC}}
\newcommand{\PDPIC}{\mathcal{P}_\text{D}^\text{IC}}
\newcommand{\PSPIC}{\mathcal{P}_\text{S}^\text{IC}}
\newcommand{\PCPIC}{\mathcal{P}_\text{C}^\text{IC}}
\newcommand{\JRDCCPPIC}{\mathcal{D}^\text{IC}}
\newcommand{\JRSCCPPIC}{\mathcal{S}^\text{IC}}
\newcommand{\JRDCCPPICmax}{\mathcal{D}_\text{M}^\text{IC}}
\newcommand{\JRDCCPPICmin}{\mathcal{D}_\text{m}^\text{IC}}
\newcommand{\JRSCCPPICmax}{\mathcal{S}_\text{M}^\text{IC}}
\newcommand{\JRSCCPPICmin}{\mathcal{S}_\text{m}^\text{IC}}
\newcommand{\PDifC}{\mathcal{P}_{\text{D}|\text{C}}}
\newcommand{\PSifC}{\mathcal{P}_{\text{S}|\text{C}}}
\newcommand{\PCifD}{\mathcal{P}_{\text{C}|\text{D}}}
\newcommand{\PCifS}{\mathcal{P}_{\text{C}|\text{S}}}
\newcommand{\rateifD}{\bar{\mathcal{R}}_{|\text{D}}}
\newcommand{\rateifS}{\bar{\mathcal{R}}_{|\text{S}}}
\newcommand{\PDifCPIC}{\mathcal{P}_{\text{D}|\text{C}}^\text{IC}}
\newcommand{\PSifCPIC}{\mathcal{P}_{\text{S}|\text{C}}^\text{IC}}
\newcommand{\PCifDPIC}{\mathcal{P}_{\text{C}|\text{D}}^\text{IC}}
\newcommand{\PCifSPIC}{\mathcal{P}_{\text{C}|\text{S}}^\text{IC}}

\newcommand{\setPC}{{\Xi_{C}}}
\newcommand{\setPFA}{{\Xi_{F}}}
\newcommand{\setPD}{{\Xi_{D}}}
\newcommand{\setPS}{{\Xi_{S}}}
\newcommand{\funcPC}{{\xi_{C}}}
\newcommand{\funcPFA}{{\xi_{F}}}
\newcommand{\funcPD}{{\xi_{D}}}
\newcommand{\funcPS}{{\xi_{S}}}

\newcommand{\setJRDCCPA}{{\Omega_{\text{D}1}}}
\newcommand{\setJRDCCPB}{{\Omega_{\text{D}2}}}
\newcommand{\setJRSCCPA}{{\Omega_{\text{S}1}}}
\newcommand{\setJRSCCPB}{{\Omega_{\text{S}2}}}
\newcommand{\funcJRDCCPA}{{\psi_{\text{D}1}}}
\newcommand{\funcJRDCCPB}{{\psi_{\text{D}2}}}
\newcommand{\funcJRSCCPA}{{\psi_{\text{S}1}}}
\newcommand{\funcJRSCCPB}{{\psi_{\text{S}2}}}

\newcommand{\setPCPIC}{{\Xi^\text{IC}_{C}}}
\newcommand{\setPFAPIC}{{\Xi^\text{IC}_{F}}}
\newcommand{\setPDPIC}{{\Xi^\text{IC}_{D}}}
\newcommand{\setPSPIC}{{\Xi^\text{IC}_{S}}}
\newcommand{\funcPCPIC}{{\xi^\text{IC}_{C}}}
\newcommand{\funcPFAPIC}{{\xi^\text{IC}_{F}}}
\newcommand{\funcPDPIC}{{\xi^\text{IC}_{D}}}
\newcommand{\funcPSPIC}{{\xi^\text{IC}_{S}}}

\newcommand{\setJRDCCPAPIC}{{\Omega^\text{IC}_{\text{D}1}}}
\newcommand{\setJRDCCPBPIC}{{\Omega^\text{IC}_{\text{D}2}}}
\newcommand{\setJRSCCPAPIC}{{\Omega^\text{IC}_{\text{S}1}}}
\newcommand{\setJRSCCPBPIC}{{\Omega^\text{IC}_{\text{S}2}}}
\newcommand{\funcJRDCCPAPIC}{{\psi^\text{IC}_{\text{D}1}}}
\newcommand{\funcJRDCCPBPIC}{{\psi^\text{IC}_{\text{D}2}}}
\newcommand{\funcJRSCCPAPIC}{{\psi^\text{IC}_{\text{S}1}}}
\newcommand{\funcJRSCCPBPIC}{{\psi^\text{IC}_{\text{S}2}}}

\newcommand{\resR}{\zeta_R}
\newcommand{\resC}{\zeta_C}
\newcommand{\res}{\zeta}

 \newacronym{JRDCCP}{JRDCCP}{Joint RADAR Detection and Communication Coverage Probability}
 \newacronym{JRSCCP}{JRSCCP}{Joint RADAR Success and Communication Coverage Probability}
 \newacronym{SINR}{SINR}{Signal to Interference plus Noise Ratio}
 \newacronym{SIR}{SIR}{Signal to Interference Ratio}
 \newacronym{DFRC}{DFRC}{Dual Function RADAR-Communication}
 \newacronym{PPP}{PPP}{Poisson Point Process}
 \newacronym{RCS}{RCS}{RADAR Cross Section}
 \newacronym{RT}{RT}{Ray Tracing}
 \newacronym{PDF}{PDF}{Probability Density Function}
 \newacronym{CDF}{CDF}{Cumulative Distribution Function}
 \newacronym{AWGN}{AWGN}{Additive White Gaussian Noise}
 \newacronym{MC}{MC}{Monte-Carlo}
 \newacronym{LT}{LT}{Laplace Transform}
 \newacronym{MIMO}{MIMO}{Multiple Input Multiple Output}
 \newacronym{TDMA}{TDMA}{Time Division Multiple Access}

\def\BibTeX{{\rm B\kern-.05em{\sc i\kern-.025em b}\kern-.08em
    T\kern-.1667em\lower.7ex\hbox{E}\kern-.125emX}}

\begin{document}

\title{Joint Performance Metrics for Integrated Sensing and Communication Systems in Automotive Scenarios}

\author{\IEEEauthorblockN{François De Saint Moulin\IEEEauthorrefmark{1}, Charles Wiame\IEEEauthorrefmark{2}, Luc Vandendorpe\IEEEauthorrefmark{3}, Claude Oestges\IEEEauthorrefmark{4}\thanks{François De Saint Moulin is a Research Fellow of the Fonds de la Recherche Scientifique - FNRS.}}\\
\IEEEauthorblockA{ICTEAM, UCLouvain - Louvain-la-Neuve, Belgium\\}
\IEEEauthorrefmark{1}francois.desaintmoulin@uclouvain.be,
\IEEEauthorrefmark{2}charles.wiame@uclouvain.be,\\
\IEEEauthorrefmark{3}luc.vandendorpe@uclouvain.be,
\IEEEauthorrefmark{4}claude.oestges@uclouvain.be}

\maketitle

\thispagestyle{plain}%to add vs delete page numbers: plain vs empty
\pagestyle{plain}

%%%%%%%%%%%%%%%%%%%%%%

\begin{abstract}
In this paper, multiple metrics are presented in order to jointly evaluate the performance of the radar and communication functions in scenarios involving Dual Function Radar Communication (DFRC) systems using stochastic geometry. These metrics are applied in an automotive scenario involving a two-lane road with vehicles and smart traffic lights, both equipped with DFRC systems. First, the performance achieved with these metrics are validated using Monte-Carlo (MC) simulations. Additionally, optimisation w.r.t. the power of the vehicles and smart traffic lights is performed based on the metrics. Then, the model is extended to include interference cancellation for the radar and/or communication function in all the metrics. Either perfect interference cancellation is applied, or a new model is proposed for imperfect interference cancellation.
\end{abstract}


%%%%%%%%%%%%%%%%%%%%%%

\section{Introduction}
In recent years, with the spectrum scarcity and the convergence of the carrier frequency ranges used by radar and communication systems, the possibility to operate both functions using the same resources has received an increased attention. Joint radar-communication systems enable to perform both functions simultaneously with a single hardware platform \cite{Zheng2019,Liu2020,Luong2021}. Two approaches have been adopted: coexistence and co-design. In the former, different waveforms are used for each function, and interference between both functions must be mitigated. In the latter, a single waveform is transmitted, and trade-offs are made in the waveform design between both functions. These are also often named Dual Function radar-Communication (\acrshort{DFRC}) systems. In order to evaluate the performance of such systems in a given scenario, multiple metrics have been defined for the radar and communication functions. One tool enabling to evaluate these metrics is stochastic geometry. \smallskip

\subsection{Related Works}

Stochastic geometry is a branch of mathematics which provides mathematical models and statistical methods to study and analyse random spatial distributions, mainly based on point processes. This tool has been widely used to assess the average performance achieved in wireless networks \cite{baccelliV1,baccelliV2,Elsawy2013,Haenggi2009,Hmamouche2021}. As a non exhaustive list, several performance metrics evaluated with this tool are the coverage probability, spectral efficiency, energy efficiency and total received power. The joint evaluation of multiple metrics has also been considered, i.e. the achieved rate in uplink and downlink \cite{Singh2015}, or the simultaneous wireless information and power transfer \cite{Renzo2016}. For radar applications, it has also been used to evaluate radar metrics (i.e. the false alarm, detection and success probability) and the impact of multiple parameters in different scenarios. For instance, \cite{Munari2018} analyses the Cumulative Distribution Function (\acrshort{CDF}) of the interference with a population of pulsed-radar devices, \cite{Ram2020,Ram2021,Ram2022} consider the radar performance in a cluttered environment, and \cite{Park2018} evaluates the detection probability when multiple obstacles are distributed around the radar systems. Considering joint radar and communication scenarios, some works have been published regarding either coexisting radar and communication or co-designed systems. In the first case, \cite{Ren2019,Ren2021} evaluate the radar range and the communication success probability of a joint radar-communication network with Time Division Multiple Access (\acrshort{TDMA}). The interference from a rotating radar sharing the same spectrum as a cellular system is considered in \cite{Kafafy2021}, while \cite{Rao2021} analyses the interference of a massive Multiple Input Multiple Output (\acrshort{MIMO}) cellular network on a radar system. Finally, the development of cooperative multi-point detection is presented in \cite{Skouroumounis2020,Skouroumounis2021}. In the second case with co-designed systems, \cite{Fang2019} evaluates the cooperative detection volume of a joint radar and communication system where targets are detected with the main beam and different stations cooperate by communicating through the secondary beams. The increase of the communication throughput with joint radar-communication system is evaluated in \cite{Ram2022_2}, and an analytical framework to optimise the system parameters is proposed.

Many automotive scenarios have been studied in a stochastic geometry framework. First, for radar applications only, \cite{Fang2020_2} compares the performance obtained with different random Radar Cross Section (\acrshort{RCS}) models. For the position of the vehicles, instead of using Poisson Point Processes (\acrshort{PPP}), \cite{Al-Hourani2018} also proposes to use a one-dimensional lattice, and \cite{Mishra2020} uses Matérn hard-core processes in two dimensions. In \cite{Fang2020}, an arbitrary number of lanes in both directions is considered, and the reflected interference from vehicles on neighbour lanes is taken into account. In \cite{Chu2020}, a multiple lane scenario is also considered, with front- and side-mounted radars with directional antenna patterns. To obtain a fine-grained analysis, \cite{Ghatak2022,Ghatak2022_2} evaluate the meta distribution of the Signal to Interference plus Noise Ratio (\acrshort{SINR}), which enables to analyse the reliability of the detection at each individual vehicle. Regarding joint radar and communication applications, the cooperative detection range is evaluated in \cite{Ma2019} and \cite{Cooperative2021}, respectively with spectrum allocation between both functions or with a joint radar-communication system. This range is defined as the total range in which targets are detected by multiple vehicles communicating with each other. \smallskip

To our best knowledge, the performance of both radar and communication functions in a single \acrshort{DFRC} system have not been investigated in a joint manner. This is especially useful when the radar and communication potentially interfere with each other, requiring a joint evaluation of the performance achieved by both functions. In this paper, we consider a two-lane road automotive scenario including crossing vehicles and smart traffic lights, both equipped with DFRC systems. In that scenario, the communication signal from the traffic light and the radar echo interfere with each other. Therefore, we design multiple metrics in order to evaluate separately or simultaneously the performance of both functions, highlight possible trade-offs and optimise multiple parameters as the vehicles and traffic lights powers, or the traffic light density.

\subsection{Contributions}
The contributions of this paper are summarised as follows:
\begin{itemize}
    \item A new automotive scenario involving smart traffic lights and vehicles equipped with joint radar-communication systems is studied within the stochastic geometry framework. A focus is made on the performance achieved for a typical vehicle receiving a communication signal from the traffic lights while simultaneously detecting the next vehicle in the same lane.
    \item In that scenario, multiple common metrics are computed, namely
    \begin{itemize}
    	\item the \textit{communication coverage probability}, i.e. the probability for the communication \acrshort{SINR} to reach a given requirement;
    	\item the \textit{radar false alarm probability}, i.e. the probability for the radar receiver to falsely detect a target which doesn't exist;
    	\item the \textit{radar detection probability}, i.e. the probability for the radar receiver to detect a real target;
    	\item the \textit{radar success probability}, i.e. the probability for the radar \acrshort{SINR} to reach a given requirement. 
    \end{itemize}
    \item Two new joint metrics are designed to evaluate the performance for both radar and communication functions, namely
    \begin{itemize}
    \item the \textit{Joint Radar Detection and Communication Coverage Probability} (\acrshort{JRDCCP}), i.e. the probability to detect a target with a given false alarm probability at the radar function, while achieving a sufficient \acrshort{SINR} at the communication function;
    \item the \textit{Joint Radar Success and Communication Coverage Probability} (\acrshort{JRSCCP}), i.e. the probability to achieve simultaneously a sufficient \acrshort{SINR} at both radar and communication functions.
\end{itemize}
Additionally, power optimisation for both vehicles and traffic lights is performed.
	\item Upper and lower bounds are proposed for the joint metrics, and conditional metrics are also defined, namely
	\begin{itemize}
		\item the coverage probability for vehicles already achieving a given detection or success probability;
		\item the detection or success probability for vehicles already achieving a given coverage probability.
	\end{itemize}
    \item A new tractable model is proposed to integrate perfect or partial interference cancellation at the radar and communication functions when they cooperate with each other. All proposed metrics are extended to include the latter.
\end{itemize}

%\subsection{Notations}

\subsection{Structure of the paper}
The paper is structured as follows: first, the system model is detailed in Section \ref{sec:system_model}. Next, multiple metrics are presented and developed in Section \ref{sec:metrics_development}. Then, numerical analysis is performed in Section \ref{sec:numerical_analysis}. Finally, the metrics from Section \ref{sec:metrics_development} are extended in Section \ref{sec:interference_cancellation} to include interference cancellation.

\section{System Model}
\label{sec:system_model}
\subsection{Considered scenario}

\begin{figure}
    \centering
    \includegraphics[width=0.9\linewidth]{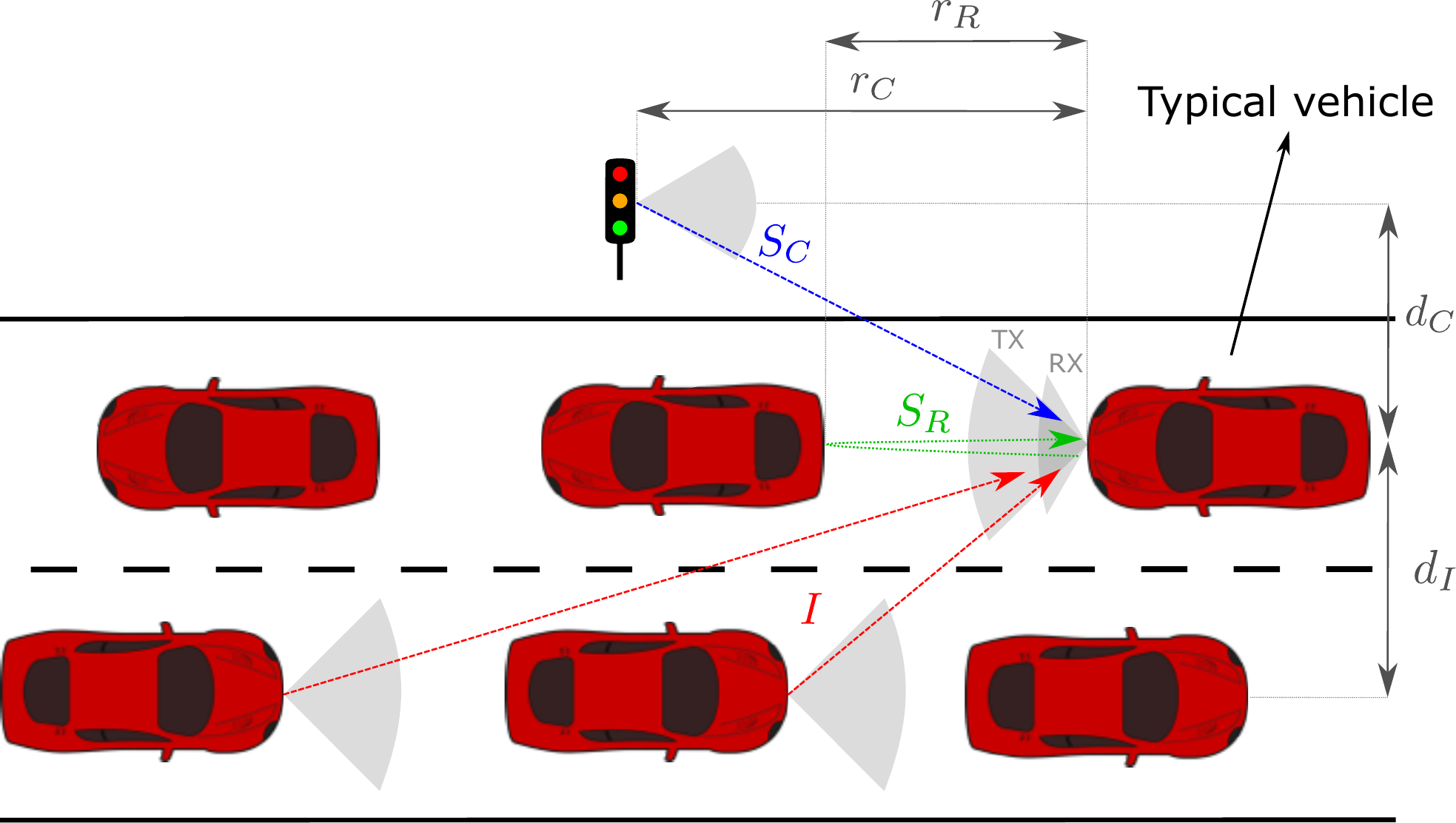}
    \caption{Considered two-lane road scenario.}
    \label{fig:automotive_scenario}
\end{figure}

We consider a two-lane road scenario, as illustrated in Figure \ref{fig:automotive_scenario}. The typical vehicle is equipped at the front side with a \acrshort{DFRC} system. Their transmit power, transmit antenna beamwidth and receive antenna beamwidth are respectively denoted as $\Pv$, $\psiVT$ and $\psiVR$. The typical vehicle receives a radar echo coming from the first vehicle driving in the same lane, and a communication signal from the nearest traffic light ahead. The transmit power and transmit antenna beamwidth of this light are respectively denoted as $\Pl$ and $\psiLT$. The vehicles driving in the opposite direction are interfering with the radar echo and communication signal. The distance perpendicular to the road direction between the typical vehicle and the smart traffic light (resp. the center of the opposite lane) is denoted $\dC$ (resp. $\dI$).

Vehicles driving in the same way, interfering vehicles on the opposite way, and traffic lights are distributed on three parallel lines following three independent \acrshort{PPP}s $\PhiR$, $\PhiI$ and $\PhiC$ of intensities respectively given by
\begin{equation}
\begin{alignedat}{10}
    &\lambdaR(r) &&=\: &\lambdaV\: &\ones{\rRmin \leq r \leq \rRmax}, \\
    &\lambdaI(r) &&=\: &\lambdaI\: &\ones{\rImin \leq r}, \\
    &\lambdaC(r) &&=\: &\lambdaL\: &\ones{\rCmin \leq r},
\end{alignedat}
\label{eq:densities}
\end{equation}
where $\lambdaV$, $\lambdaI$ and $\lambdaL$ are respectively the densities of vehicles driving on the same lane as the typical vehicle, interfering vehicles and traffic lights. The minimum and maximum detectable range of the radar function, the minimum distance (parallel to the road) of the communication signal and the minimum distance (parallel to the road) of the interfering signals are respectively denoted by $\rRmin$, $\rRmax$, $\rCmin$ and $\rImin$. Note that $\lambdaI \neq \lambdaV$ since we assume that the vehicles in the opposite lane are not always interfering: either the different vehicles transmit asynchronously, or they are not all equipped with \acrshort{DFRC} systems.  Additionally, $\rCmin$ and $\rImin$ are directly related to the antenna beamwidths of the different systems: $\rCmin = \dC/\tan(\min(\psiLT,\psiVR)/2)$ and $\rImin = \dI/\tan(\min(\psiVT,\psiVR)/2)$.

\subsection{Propagation models}
Let us consider a linear path-loss model $[\PLF_k]_\dB = [\FI_k]_\dB + 10\PLE_k\:\log_{10}(r)$, where $k=\{R,C,I\}$ designates the considered link, $\PLE_k$ is the path-loss exponent and $\FI_k$ is the intercept. The received power for the communication link is given by the Friis equation \cite{Friis} as 
\begin{equation}
    \Sc(r) = \Pl \Gc \frac{c^2}{4\pi \fc^2} \PLF_C^{-1}\left(\sqrt{r^2 +  \dC^2}\right) = \rhoC \left(r^2 + \dC^2\right)^{-\frac{\PLE_C}{2}}
    \label{eq:Sc}
\end{equation}
at a distance $r$ (parallel to the road), with $\rhoC = \frac{\Pl \Gc c^2}{4\pi \fc^2 \FI_C}$. The speed of light is denoted by $c$, and the carrier frequency by $\fc$, and $\Gc$ is the beamforming gain of the communication link. The interfering power is obtained in the same manner:
\begin{align}
    \nonumber \I &= \sum_{i\:|\:\phi_i \in \PhiI} \Pv \Gi \: \frac{c^2}{4\pi \fc^2} \: \PLF_I^{-1}\left(\sqrt{\rVert\phi_i\rVert^2 + \dI^2}\right)|h_i|^2 \\ 
    &= \sum_{i\:|\:\phi_i \in \PhiI} \rhoI \left(\rVert\phi_i\rVert^2 + \dI^2\right)^{-\frac{\PLE_I}{2}} |h_i|^2,
\end{align}
where $\rhoI = \frac{\Pv\Gi c^2}{4\pi\fc^2\FI_I}$. The beamforming gain of the interfering links is denoted by $\Gi$, and $|h_i|^2$ are i.i.d. Rician distributed small-scale fading random variables. Note that the small-scale fading has been neglected for the communication link for tractability. However, it is supposed to be highly rician since the traffic light is located at the same side of the road (and not in the traffic as the interfering links), and high frequencies (24/77\GHz) are considered for this application.
\smallskip

In the literature, the received power for the radar link is usually given by the radar equation \cite{radar_fund}:
\begin{equation}
        \SrRCS(r) = \Pv\Gr\:\frac{c^2}{4\pi \fc^2}\:\PLF_R^{-2}(r)\:\RCS
        \label{eq:SrRCS}
\end{equation}
    with a distance $r$ between the radar and the target, where $\sigma$ is the RCS of the target, and $\Gr$ is the beamforming gain of the radar link. However, the RCS varies with the distance to the target, its geometry and the considered frequencies. The issue is often tackled by modelling the RCS as an exponentially distributed random variable, following the Swerling I model. Additionally, this model is only valid in far field, i.e. when $r \geq \frac{2D^2 f_c}{c}$ with $D$ the largest dimension of the reflector. For instance, at a distance of $5$m, it requires reflectors with largest dimensions $D \leq \sqrt{\frac{rc}{2f_c}} \approx 18$cm or $10$cm respectively at carrier frequencies of $24$ and $77$GHz. To alleviate the issues inherent to the RCS-based model, we instead modelled the received power for the radar link as
\begin{equation}
        \Sr(r) = \Pv\Gr\:\frac{c^2}{4\pi \fc^2}\:\PLF_R^{-1}(2r) = \rhoR \: r^{-\alpha_R}
        \label{eq:SrRT}
\end{equation}
with a distance $r$ between the radar and the target. This model is only valid for reflector dimensions larger than the first Fresnel zones diameter, namely $D \geq \sqrt{\frac{rc}{f_c}} \approx 25$ or 
$14$cm at a distance of $5$m, and carrier frequencies of $24$ and $77$GHz. Therefore, it is suited for the considered automotive scenario. Again, the small-scale fading is neglected for tractability, but it is also supposed to be highly rician considering the scenario and the high frequencies used in automotive radar applications. \smallskip

To simplify the notations, we define $\Sr \equiv \Sr(\rR)$ where $\rR$ is the distance between the vehicle of interest and the vehicle driving ahead, and $\Sc \equiv \Sc(\rC)$, where $\rC$ is the distance (parallel to the road) between the vehicle of interest and the nearest traffic light ahead. We also define $\Srmin =\Sr(\rRmax)$, $\Srmax = \Sr(\rRmin)$ and $\Scmax = \Sc(\rCmin)$ as the minimum and maximum possible power levels for the radar and communication links. Let us also define the following operator:
\begin{equation*}
\restesp{A}{g(\Sr,\Sc)} = \esp{\Sr,\Sc}{g(\Sr,\Sc)\cdot\ones{(\Sr,\Sc)\in A}}
\end{equation*}
which computes the expectation over $\Sr$ and $\Sc$ of a function $g$ in a given domain $A \in \mathbb{R}^2$.

\section{Development of the Metrics}
\label{sec:metrics_development}
\begin{figure}[H]
    \centering
    \includegraphics[width=0.9\linewidth]{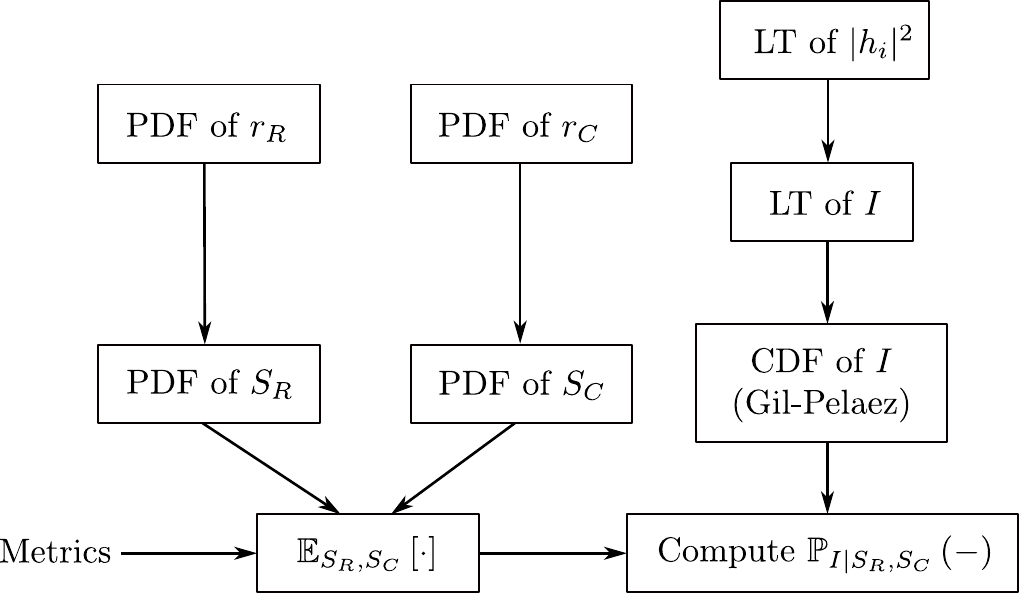}
    \caption{Metrics development steps.}
    \label{fig:metrics_development}
\end{figure}

In this section, multiple metrics are presented and developed, following the different steps of Figure \ref{fig:metrics_development}. 

First, the Probability Density Functions (\acrshort{PDF}) of $\rR$ and $\rC$ are computed. Then, the PDFs of $\Sr$ and $\Sc$ are derived from the \acrshort{PDF}s of their associated distances. This enables to condition the metrics over the powers of the useful signals, and develop their expressions to obtain a function of the \acrshort{CDF} of $\I$. Finally, this \acrshort{CDF} is computed based on the Laplace Transform (\acrshort{LT}) of $\I$ --- itself function of the \acrshort{LT} of the small-scale fading of the interfering links $|h_i|^2$ --- thanks to the Gil-Pelaez theorem \cite{GPtheorem}. \smallskip

In Sections \ref{sec:PDFrRrC} to \ref{sec:LTI}, all preliminary results needed in order to develop the metrics are computed. Next, usual radar and communication metrics, namely
\begin{enumerate}
\item the communication coverage probability;
\item the radar detection and false alarm probabilities;
\item the radar success probability,
\end{enumerate}
are presented and derived in Sections \ref{sec:PC} to \ref{sec:PS}. Then, based on these different metrics, the two joint metrics, namely
\begin{enumerate}
\item the JRDCCP metric;
\item the JRSCCP metric,
\end{enumerate} 
are designed in Sections \ref{sec:JRDCCP} and \ref{sec:JRSCCP}. Upper and lower bounds are also proposed. Finally, conditional metrics are defined in Section \ref{sec:cond_metrics} based on the given radar, communication and joint metrics.

\subsection{Probability density functions of $\rR$ and $\rC$}
\label{sec:PDFrRrC}

Based on the densities given in \eqref{eq:densities}, the \acrshort{PDF}s of $\rR$ and $\rC$ are presented in Lemma \ref{lemma:PDFr}. 

\lemma{PDFr}{Assuming that there is at least one vehicle in the radar detectable range, the \acrshort{PDF}s of $\rR$ and $\rC$ are respectively given by
{\normalfont
\begin{multline}
    \PDF{\rR}{r} = \frac{\lambdaV\:\exp{-\lambdaV(r-\rRmin)}}{1-\exp{-\lambdaV(\rRmax-\rRmin)}} \\ \cdot \ones{\rRmin \leq r \leq \rRmax}
\end{multline}
\label{eq:PDFrR}
}
and
{\normalfont
\begin{equation}
    \PDF{\rC}{r} = \lambdaL\:\exp{-\lambdaL (r-\rCmin)} \cdot\ones{\rCmin \leq r}.
    \label{eq:PDFrC}
\end{equation}
}
}{The proof is detailed in Appendix \ref{app:proof_PDF_r}.}

\subsection{Probability density functions of $\Sr$ and $\Sc$}
Based on the expressions of Lemma \ref{lemma:PDFr}, the expressions of the \acrshort{PDF}s of $\Sr$ and $\Sc$ are developed in Lemma \ref{lemma:PDFs}.

\lemma{PDFs}{Assuming that there is at least one vehicle in the radar detectable range, the \acrshort{PDF}s of $\Sr$ and $\Sc$ are respectively given by
{\normalfont
\begin{multline}
    \PDF{\Sr}{s} =  \frac{\rR(s)}{\PLE_R\:s}\frac{\lambdaV\:\exp{-\lambdaV(\rR(s)-\rRmin)}}{1-\exp{-\lambdaV(\rRmax - \rRmin)}} \\\cdot \ones{\Srmin \leq s \leq \Srmax}
    \label{eq:PDFsRrt}
\end{multline}
}
with $\rR(s) = \rhoR^{\frac{1}{\PLE_R}}\:s^{-\frac{1}{\PLE_R}}$ , and
{\normalfont
\begin{multline}
    \PDF{\Sc}{s} =  \frac{\rC^2(s) + \dC^2}{\PLE_C\:s\:\rC(s)}\:\lambdaL\: \exp{-\lambdaL(\rC(s)-\rCmin)} \\\cdot \ones{0 \leq s \leq \Scmax}
    \label{eq:PDFsC}
\end{multline}
with $\rC(s) = \sqrt{\rhoC^{\frac{2}{\PLE_C}}\:s^{-\frac{2}{\PLE_C}}-\dC^2}$.
}
}{The proof is detailed in Appendix \ref{app:proof_PDF_s}.}

\subsection{Laplace transform of the small-scale fading for the interfering links}

Under Rician fading, the Laplace transform of the small-scale fading random variables is given by Lemma \ref{lemma:LTh}.

\lemma{LTh}{The Laplace transform of the small-scale fading for the interfering links is given by
{\normalfont
\begin{equation}
    \LT_{|h_i|^2}(s) = \frac{K+1}{K+1+s}\:\exp{-\frac{K\:s}{K+1+s}},
    \label{eq:LTh}
\end{equation}
where $K$ is the Rician K-factor of the interfering links.
}
}{The proof is detailed in Appendix \ref{app:proof_LT_h}.}

\subsection{Laplace transform of the interfering power $\I$}
\label{sec:LTI}
Based on Lemma \ref{lemma:LTh}, the expression of the Laplace transform of the interfering power $\I$ is obtained in Lemma \ref{lemma:LTint}.

\lemma{LTint}{The Laplace transform of the interfering power $\I$ is given by
{\normalfont
\begin{align}
&\LT_\I(s) \\
\nonumber \quad &=\exp{-\lambdaI\int_{\rImin}^\infty \left(1-\LT_{|h_i|^2}\left(s \rhoI \left(r^2 + \dI^2\right)^{-\frac{\PLE_I}{2}}\right)\right)\d r},
\label{eq:LTint}
\end{align}
}
where $\LT_{|h_i|^2}$ denotes the Laplace transform of the small-scale fading for the interfering links.
}{
The proof is detailed in Appendix \ref{app:proof_LT_I}.
}

\subsection{Coverage probability}
\label{sec:PC}

The \textit{coverage probability} is defined as the probability to achieve a sufficient \acrshort{SINR} for the communication function. Additionally, ensuring that the \acrshort{SINR} is higher than $\threshC$ also ensures that the achievable communication rate is higher than $\frac{1}{T_c}\log_2(1+\threshC)$ where $1/T_c$ is the symbol rate of the transmission. \smallskip

In the considered scenario, the radar echo from the next vehicle ahead and the vehicles on the opposite lane are interfering with the communication function. Assuming that the scenario is interference-limited, this metric is written as
\begin{equation}
\PC(\threshC) = \Pr{}{\frac{\Sc}{\Sr + \I} \geq \threshC}.
\end{equation}
Proposition \ref{prop:PC} gives the expression of the coverage probability depending on the required communication Signal to Interference Ratio (\acrshort{SIR}) threshold. 

\proposition{PC}{For a given communication \acrshort{SIR} threshold $\threshC$, the coverage probability is obtained as
{\normalfont
\begin{equation}
\PC(\threshC) = \restesp{\setPC}{\frac{1}{2}-\frac{1}{\pi}\int_0^\infty \frac{1}{\tau}\:\Im{\LT_\I(-j\tau) \: \funcPC(\tau)} \:\mathrm{d}\tau}
\label{eq:CP_eq}
\end{equation}
}
where
{\normalfont
\begin{equation}
\funcPC(\tau) = \exp{-j\tau\left(\frac{\Sc}{\threshC} - \Sr\right)}
\label{eq:CP_func}
\end{equation}
}
and the area $\setPC$ is defined as
{\normalfont
\begin{equation}
\setPC \triangleq \left\{
\begin{alignedat}{10}
&\Srmin \leq \Sr \leq \Srmax \\
&\threshC \Sr \leq \Sc \leq \Scmax
\end{alignedat}
\right\}.
\label{eq:CP_set}
\end{equation}
}
}{The proof is detailed in Appendix \ref{app:proof_CP}.}

Note that the \acrshort{CDF} of the \textit{spectral efficiency} can be computed based on the coverage probability:
\begin{equation}
\rateCCDF(\eta) = \Pr{}{\log_2\left(1 + \frac{\Sc}{\Sr + I}\right) \geq \eta} = \PC\left(2^{\eta}-1\right).
\end{equation}
Similarly, the average spectral efficiency is computed as
\begin{equation}
\rate = \int_0^\infty \rateCCDF(\eta) \:\d \eta = \frac{1}{\ln 2} \int_0^\infty \frac{\PC(\threshC)}{1+\threshC} \:\d \threshC,
\label{eq:rate}
\end{equation}
applying the change of variable $\theta = 2^\eta - 1$.

\subsection{Detection and false alarm probabilities}
\label{sec:PD_PFA}
Let us consider the output of the radar receiver before the detector. As done in \cite{radar_fund}, two hypotheses are defined for each cell, i.e. each element of the delay-Doppler-angle map produced at its output: 
\begin{itemize}
    \item[-] $H_0$ : there is no target within the cell;
    \item[-] $H_1$ : there is a target within the cell. 
\end{itemize}
The amplitude of the cell under test is described as
\begin{equation}
    x = \left\{
    \begin{alignedat}{10}
    & A + w\quad && \text{under } H_1,\\
    & w && \text{under } H_0,
    \end{alignedat}
    \right.
\end{equation}
where $|A|^2 = \Sr \RPG$ with a radar processing gain $\RPG$, $w$ is an Additive White Gaussian Noise (\acrshort{AWGN}) of variance $\sigma_w^2$ modelling the noise and the interference. For a given threshold $\threshRd$, the \textit{false alarm probability} $\PFA$ and \textit{detection probability} $\PD$ are respectively defined as 
\begin{alignat}{10}
&\PFAcut(\threshRd) &&= \Pr{x}{|x|^2 \geq \threshRd\:|\:H_0}, \\
&\PDcut(\threshRd) &&= \Pr{x}{|x|^2 \geq \threshRd\:|\:H_1}.
\end{alignat}

In practice, a given false alarm probability is required, and the threshold is defined to meet this requirement.\smallskip

In the considered automotive scenario, the nearest traffic light ahead and the vehicles on the opposite lane are interfering with the radar function. Assuming that the scenario is interference-limited, these metrics are respectively rewritten as follows:
\begin{alignat}{10}
&\PFA(\threshRd) &&= \Pr{\Sc,\I}{\Sc+\I\geq \threshRd}, \\
&\PD(\threshRd) &&= \Pr{}{\Sr\RPG + \Sc + \I \geq \threshRd}.
\label{eq:PD_PFA}
\end{alignat}
Propositions \ref{prop:PFA} and \ref{prop:PD} provide expressions of the false alarm and detection probabilities depending on the threshold of the radar detector.

\proposition{PFA}{For a given threshold $\threshRd$, the false alarm probability is given by
{\normalfont
\begin{align}
    \PFA(\threshRd) &= 1 - \restesp{\setPFA}{\frac{1}{2} - \frac{1}{\pi}\int_0^\infty \frac{1}{\tau}\:\Im{\LT_\I(-j\tau)\:\funcPFA(\tau,s)}\d \tau} 
    \label{eq:FAP_eq}
\end{align}
}
where
{\normalfont
\begin{equation}
    \funcPFA(\tau) = \exp{-j\tau(\threshRd - \Sc)},
    \label{eq:FAP_func}
\end{equation}
}
and the area $\setPFA$ is defined as
{\normalfont
\begin{equation}
\setPFA \triangleq \left\{
\begin{alignedat}{10}
&\Sc \leq \threshRd \\
&0 \leq \Sc \leq \Scmax
\end{alignedat}
\right\}.
\label{eq:FAP_set}
\end{equation}
}
}{The proof is detailed in Appendix \ref{app:proof_FAP}.}

\proposition{PD}{For a given threshold $\threshRd$, the detection probability is given by
{\normalfont
\begin{align}
     &\PD(\threshRd) = 1 - \restesp{\setPD}{\frac{1}{2} - \frac{1}{\pi}\int_0^\infty \frac{1}{\tau}\:\Im{\LT_\I(-j\tau)\:\funcPD(\tau)}\d\tau}
    \label{eq:DP_eq}
\end{align} 
}
where
{\normalfont
\begin{equation}
    \funcPD(\tau) = \exp{-j\tau(\threshRd - \Sr\RPG - \Sc)},
    \label{eq:DP_func}
\end{equation}
}
and the area $\setPD$ is defined as
\begin{equation}
\setPD \triangleq \left\{
\begin{alignedat}{10}
&\Sc \leq \threshRd - \Sr\RPG \\
&\Srmin \leq \Sr \leq \Srmax \\
&0 \leq \Sc \leq \Scmax
\end{alignedat}
\right\}.
\label{eq:DP_set}
\end{equation}
}{The proof is detailed in Appendix \ref{app:proof_DP}.}

\subsection{Success probability}
\label{sec:PS}
Following the hypothesis testing problem presented in Section \ref{sec:PD_PFA}, for a given false alarm probability $\PFA$, the detection probability is given by \cite{radar_fund}
\begin{equation}
    \PDcut = Q_1\left(\sqrt{2 |A|^2/\sigma_w^2}, \sqrt{-2\ln\PFA}\right),
\end{equation}
where $Q_1$ is the first order Marcum Q-function. Therefore, since an increase of the radar SINR leads to a higher detection probability, the \textit{success probability} has been defined as
\begin{equation}
    \PScut(\threshRs) = \Pr{}{|A|^2/\sigma_w^2 \geq \threshRs}
\end{equation}
where $\threshRs$ is a given radar SINR threshold.\smallskip

In the considered automotive scenario, the nearest traffic light ahead and the vehicles on the opposite lane interfere with the radar function. Assuming that the scenario is interference-limited, this metric is rewritten as follows:
\begin{equation}
\PS(\threshRs) =  \Pr{}{\frac{\Sr \RPG}{\Sc + \I} \geq \threshRs}.
\end{equation}
Proposition \ref{prop:PS} gives the expression of the success probability depending on the required radar \acrshort{SIR} threshold.

\proposition{PS}{For a given radar \acrshort{SIR} threshold $\threshRs$, the success probability is obtained as
{\normalfont
\begin{align}
&\PS(\threshRs) = \restesp{\setPS}{\frac{1}{2} - \frac{1}{\pi}\int_0^\infty \frac{1}{\tau}\:\Im{\LT_\I(-j\tau)\:\funcPS(\tau)}\:\mathrm{d}\tau}
\label{eq:SP_eq}
\end{align}
}
where
{\normalfont
\begin{equation}
\funcPS(\tau) = \exp{-j\tau\left(\frac{\Sr\RPG}{\threshRs} - \Sc\right)},
\label{eq:SP_func}
\end{equation}
}
and the area $\setPS$ is defined as
\begin{equation}
\setPS \triangleq \left\{
\begin{alignedat}{10}
&\Sc \leq \Sr\RPG/\threshRs \\
&\Srmin \leq \Sr \leq \Srmax \\
&0 \leq \Sc \leq \Scmax
\end{alignedat}
\right\}.
\label{eq:SP_set}
\end{equation}
}{The proof is detailed in Appendix \ref{app:proof_SP}.}

\subsection{\acrshort{JRDCCP} metric}
\label{sec:JRDCCP}

We define the \acrshort{JRDCCP} metric as an extension of the communication coverage and the radar detection probabilities for radar-communication scenarios:
\begin{align}
    &\JRDCCP(\threshC;\gamma) = \Pr{}{\Sr\RPG + \Sc + \I \geq \threshRd, \: \frac{\Sc}{\Sr + \I} \geq \threshC} \label{eq:JRDCCP_def}\\
    \nonumber &\text{for a given } \PFA(\threshRd).
\end{align}
This gives the probability that a radar detection occurs for a given false alarm probability, while ensuring a sufficient \acrshort{SIR} for the communication function. The threshold $\threshRd$ is selected in order to fulfill the false alarm probability requirement, and the \acrshort{JRDCCP} metric is then obtained for a given communication \acrshort{SIR} threshold $\threshC$. The metric is developed in Proposition \ref{prop:JRDCCP}.

\proposition{JRDCCP}{
For a given false alarm probability $\PFA$ (and therefore for a given threshold $\threshRd$) and a given communication SIR threshold $\threshC$, the \acrshort{JRDCCP} metric is obtained as
{\normalfont
\begin{align}
\nonumber &\JRDCCP(\threshC;\gamma) = -\restesp{\setJRDCCPB}{\frac{1}{\pi}\int_0^\infty  \frac{1}{\tau}\:\Im{\LT_\I(-j\tau)\:\funcJRDCCPB(\tau)}\d\tau} \\
& + \restesp{\setJRDCCPA}{\frac{1}{2} - \frac{1}{\pi} \int_0^\infty\frac{1}{\tau}\:\Im{\LT_\I(-j\tau)\: \funcJRDCCPA(\tau)}\d\tau} \label{eq:JRDCCP_eq}
\end{align}
}
where
{\normalfont
\begin{align}
    \funcJRDCCPA(\tau) &= \funcPC(\tau), \quad \funcJRDCCPB(\tau) = \funcPC(\tau) - \funcPD(\tau),
    \label{eq:JRDCCP_func}
\end{align}
}
and the areas $\setJRDCCPA$ and $\setJRDCCPB$ are defined as
{\normalfont
\begin{align}
    \setJRDCCPA &\triangleq \left\{
    \begin{alignedat}{10}
    &\Sc \leq \max\left(\threshC\Sr, \threshRd - \Sr \RPG\right) \\
    &\Srmin \leq \Sr \leq \Srmax \\
    &0 \leq \Sc \leq \Scmax
    \end{alignedat}
    \right\}, \\
    \setJRDCCPB &\triangleq \left\{
    \begin{alignedat}{10}
    &\frac{\threshC\threshRd}{\threshC+1} - \Sr \frac{\threshC(\RPG-1)}{\threshC + 1} \leq \Sc \leq \threshRd - \RPG \Sr \\
    &\Srmin \leq \Sr \leq \Srmax \\
    &0 \leq \Sc \leq \Scmax
    \end{alignedat}
    \right\}.
    \label{eq:JRDCCP_set}
\end{align}
}
}{The proof is detailed in Appendix \ref{app:proof_JRDCCP}.}

\subsection{\acrshort{JRSCCP} metric}
\label{sec:JRSCCP}
We define the \acrshort{JRSCCP} metric as an extension of the success probability for radar-communication scenarios:
\begin{equation}
    \JRSCCP(\threshRs,\threshC) = \Pr{}{\frac{\Sr\RPG}{\Sc + \I} \geq \threshRs, \: \frac{\Sc}{\Sr + \I} \geq \threshC}.
    \label{eq:JRSCCP_def}
\end{equation}
It gives the probability that both the radar and communication \acrshort{SIR}s are sufficient compared to the threshold requirements. This metric is developed in Proposition \ref{prop:JRSCCP}.

\proposition{JRSCCP}{For a given radar \acrshort{SIR} threshold $\threshRs$ and communication \acrshort{SIR} threshold $\threshC$, if the following condition is fulfilled:
{\normalfont
\begin{equation}
[\threshRs]_{\dB} + [\threshC]_{\dB} \leq [\RPG]_{\dB},
\label{eq:JRSCCP_cond}
\end{equation}
} 
the \acrshort{JRSCCP} metric is obtained as
{\normalfont
\begin{align}
   \nonumber &\JRSCCP(\threshRs,\threshC) = \restesp{\setJRSCCPA}{\frac{1}{2} - \frac{1}{\pi} \int_0^\infty \frac{1}{\tau}\:\Im{\LT_\I(-j\tau) \: \funcJRSCCPA(\tau)} \d \tau} \\
     +& \restesp{\setJRSCCPB}{\frac{1}{2}-\frac{1}{\pi} \int_0^\infty \frac{1}{\tau}\:\Im{\LT_\I(-j\tau) \: \funcJRSCCPB(\tau)} \d \tau} 
     \label{eq:JRSCCP_eq}
\end{align}
}
where 
{\normalfont
\begin{align}
    \funcJRSCCPA(\tau) &= \funcPS(\tau),\quad \funcJRSCCPB(\tau) = \funcPC(\tau),
    \label{eq:JRSCCP_func}
\end{align}
}
and the areas $\setJRSCCPA$ and $\setJRSCCPB$ are defined as
{\normalfont
\begin{align}
    \setJRSCCPA &\triangleq \left\{
    \begin{alignedat}{10}
    &\Sr \frac{\threshC(\threshRs+\RPG)}{\threshRs(\threshC+1)} \leq \Sc \leq \frac{\Sr\RPG}{\threshRs} \\
    &\Srmin \leq \Sr \leq \Srmax \\
    &0 \leq \Sc \leq \Scmax
    \end{alignedat}
    \right\}, \\
    \setJRSCCPB &\triangleq \left\{
    \begin{alignedat}{10}
    & \threshC\Sr \leq \Sc \leq \Sr \frac{\threshC(\threshRs+\RPG)}{\threshRs(\threshC+1)}\\
    &\Srmin \leq \Sr \leq \Srmax \\
    &0 \leq \Sc \leq \Scmax
    \end{alignedat}
    \right\}.
    \label{eq:JRSCCP_set}
\end{align}
}
Otherwise, the metric is equal to zero.
}{The proof is detailed in Appendix \ref{app:proof_JRSCCP}.}

Using Fréchet inequalities \cite{Frechet1935}, upper and lower bounds are proposed for the \acrshort{JRDCCP} and \acrshort{JRSCCP} metrics in Corrolary \ref{corr:JRDSCCP_bounds}.

\corrolary{JRDSCCP_bounds}
{
For a given false alarm probability $\PFA$ (and therefore for a given threshold $\threshRd$) and a given communication SIR threshold $\threshC$, the JRDCCP metric is bounded by
{\normalfont
\begin{equation}
\JRDCCPmin(\threshC;\threshRd) \leq \JRDCCP(\threshC;\threshRd) \leq \JRDCCPmax(\threshC;\threshRd),
\end{equation}
}
with
{\normalfont
\begin{align}
\JRDCCPmin(\threshC;\threshRd) &= \max\left(0,\PD(\gamma) + \PC(\threshC) -1 \right), \\
\JRDCCPmax(\threshC;\threshRd) &= \min\left(\PD(\threshRd),\PC(\threshC)\right).
\end{align}
}
Similarly, for a given radar \acrshort{SIR} threshold $\threshRs$ and communication \acrshort{SIR} threshold $\threshC$, the JRSCCP metric is bounded by
{\normalfont
\begin{equation}
\JRSCCPmin(\threshRs,\threshC) \leq \JRSCCP(\threshRs,\threshC) \leq \JRSCCPmax(\threshRs,\threshC),
\end{equation}
}
with
{\normalfont
\begin{align}
\JRSCCPmin(\threshRs,\threshC) &= \max\left(0,\PS(\threshRs) + \PC(\threshC) -1 \right), \\
\JRSCCPmax(\threshRs,\threshC) &= \min\left(\PS(\threshRs),\PC(\threshC)\right).
\end{align}
}
}
{
The proposed bounds are given by Fréchet inequalities.
}

\subsection{Conditional metrics}    
\label{sec:cond_metrics}
One can be interested in the performance achieved by vehicles already fulfilling successfully one of the two functions. On the one hand, for the radar function, it is supposed that either a radar detection has occured, or that the radar \acrshort{SIR} fulfills a given threshold requirement. Conditional metrics are then respectively expressed as follows:
\begin{align}
\PCifD(\threshC;\threshRd) &= \Pr{}{\frac{\Sc}{\Sr + \I} \geq \threshC \:\bigg|\: \Sr\RPG + \Sc + \I \geq \threshRd} \\
    \nonumber &\text{for a given } \PFA(\threshRd) , \\
\PCifS(\threshC;\threshRs) &= \Pr{}{\frac{\Sc}{\Sr + \I} \geq \threshC \:\bigg|\: \frac{\Sr\RPG}{\Sc + \I} \geq \threshRs }.
\end{align}
On the other hand, for the communication function, it is supposed that the communication \acrshort{SIR} fulfills a given threshold requirement. Conditional metrics are then expressed as follows:
\begin{align}
\PDifC(\threshRd;\threshC) &= \Pr{}{\Sr\RPG + \Sc + \I \geq \threshRd \:\bigg|\: \frac{\Sc}{\Sr + \I} \geq \threshC} \\
    \nonumber &\text{for a given } \PFA(\threshRd) , \\
\PSifC(\threshRs;\threshC) &= \Pr{}{\frac{\Sr\RPG}{\Sc + \I} \geq \threshRs  \:\bigg|\: \frac{\Sc}{\Sr + \I} \geq \threshC}.
\end{align}
The different conditional metrics are detailed in Corrolary \ref{corr:condMetCDS}.

\corrolary{condMetCDS}{For a given false alarm probability $\PFA$ (and therefore for a given threshold $\threshRd$) or a given radar \acrshort{SIR} threshold $\threshRs$, the coverage probabilities achieved by vehicles performing successfully the radar function for a given communication \acrshort{SIR} threshold $\threshC$ are given by
{\normalfont
\begin{equation}
\PCifD(\threshC;\threshRd) = \JRDCCP(\threshC;\threshRd) / \PD(\threshRd)
\end{equation}
}
and
{\normalfont
\begin{equation}
\PCifS(\threshC;\threshRs) = \JRSCCP(\threshRs,\threshC) / \PS(\threshRs).
\end{equation}
}
Similarly, the detection and success probabilities achieved by vehicles performing successfully the communication function are given by
{\normalfont
\begin{equation}
\PDifC(\threshRd;\threshC) = \JRDCCP(\threshC;\threshRd) / \PC(\threshC)
\end{equation}
}
and
{\normalfont
\begin{equation}
\PSifC(\threshRs;\threshC) = \JRSCCP(\threshRs,\threshC) / \PC(\threshC).
\end{equation}
}
}{
These results are obtained using Bayes law.
}

Note that the average spectral efficiency can also be evaluated for these users following \eqref{eq:rate}:
\begin{align*}
\rateifD &= \frac{1}{\ln 2} \int_0^\infty \frac{\PCifD(x)}{1+x} \:\d x, \\
\rateifS &= \frac{1}{\ln 2} \int_0^\infty \frac{\PCifS(x)}{1+x} \:\d x.
\end{align*}

\section{Numerical Analysis}
\label{sec:numerical_analysis}
\begin{figure*}
\centering
\includegraphics[width=0.85\textwidth]{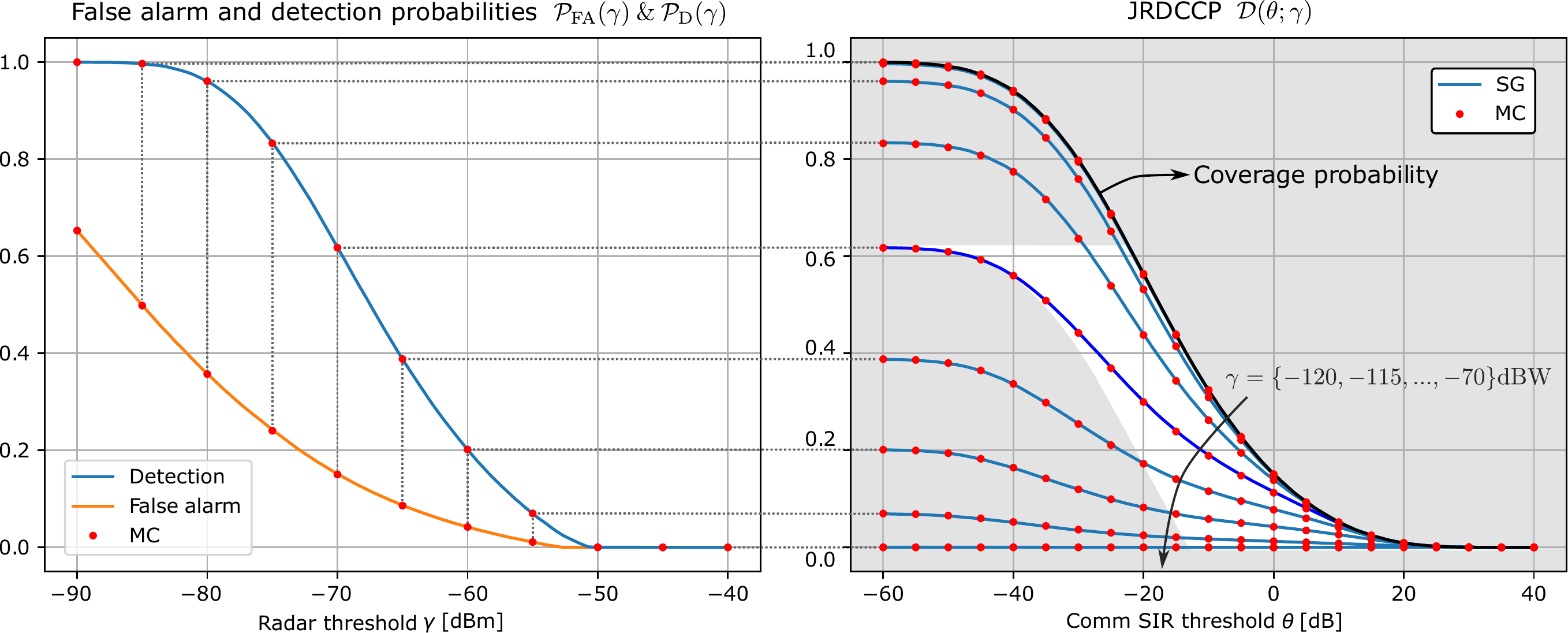}
\caption{False alarm probability, detection probability (left) and \acrshort{JRDCCP} metric (right). The white area corresponds to the region delimited by the upper and lower bounds of the magenta curve.}
\label{fig:JRDCCP}
\end{figure*}

\begin{figure*}
\centering
\includegraphics[width=0.85\textwidth]{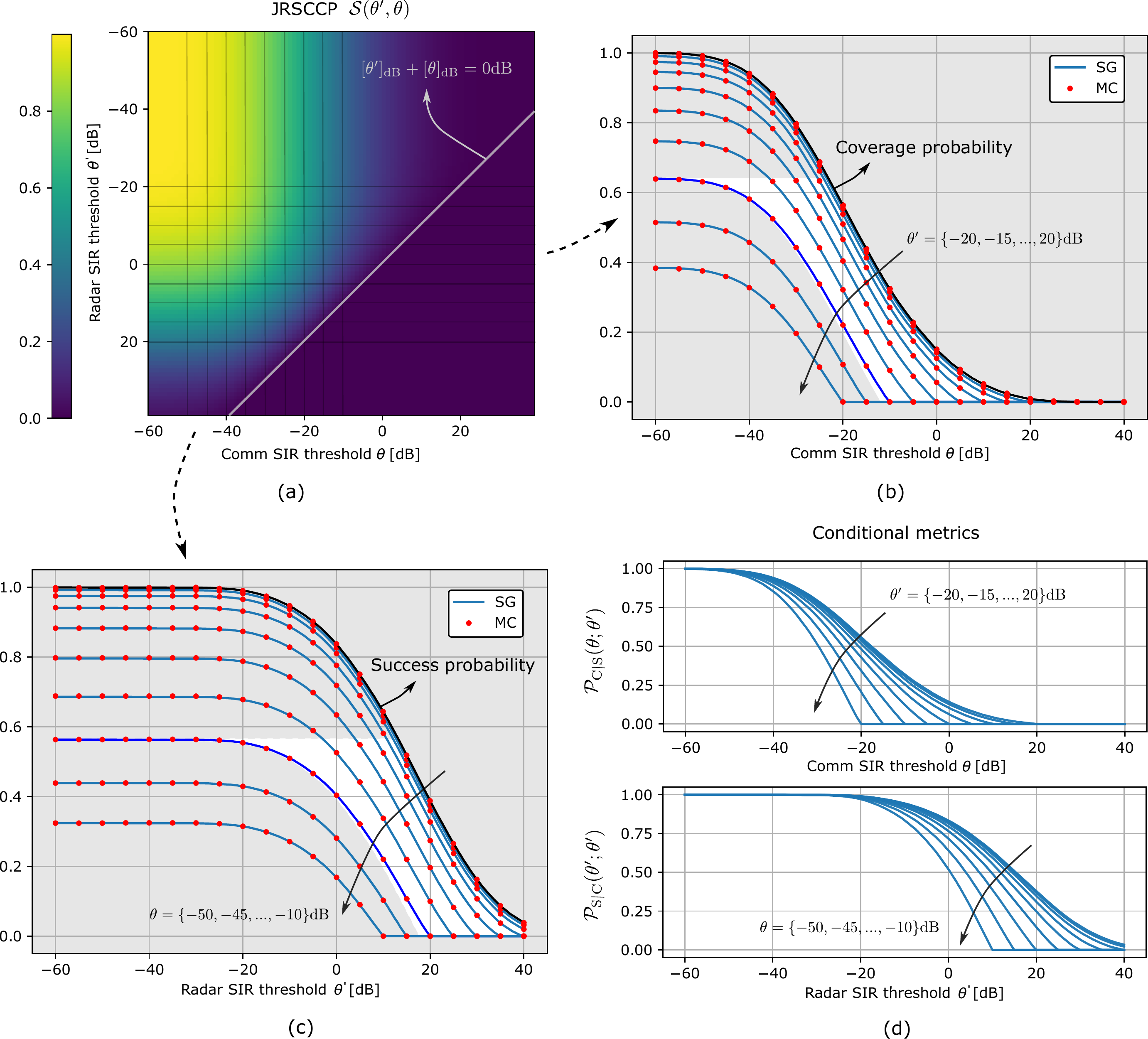}
\caption{\acrshort{JRSCCP} metric (a), cut along the $\threshC$-axis (b), cut along the $\threshRs$-axis (c) and conditional metrics (d). The white areas in (b) and (c) correspond to the regions delimited by the upper and lower bounds of the magenta curves.}
\label{fig:JRSCCP}
\end{figure*}

Table \ref{tab:scenar_params} summarises the parameters used in the simulated scenarios.

\begin{table}[H]
\centering
\caption{Scenario parameters.}
\renewcommand{\arraystretch}{1.5}
\begin{tabular}{|c|c|}
\hline
$P_V = P_L = 1$W & $G_R = G_C = G_I = 1$ \\\hline
$\kappa = 1$ & $d_C = 2.5$m, $d_I = 3$m \\\hline
\multicolumn{2}{|c|}{$\rRmin = 5$m, $\psiVT=22.5\degree$, $\psiVR=\psiLT=45\degree$} \\\hline
\multicolumn{2}{|c|}{$\lambda_V = 0.02\text{m}^{-1}$, $\lambda_L = 0.01\text{m}^{-1}$, $\lambda_I = 0.002\text{m}^{-1}$} \\\hline
$\beta_R = \beta_C = \beta_I = 4\pi$ & $\alpha_R = 2$, $\alpha_C = \alpha_I = 3$\\\hline
\end{tabular}
\renewcommand{\arraystretch}{1}
\label{tab:scenar_params}
\end{table}

\subsection{Metrics analysis}

Figure \ref{fig:JRDCCP} illustrates the false alarm probability, the detection probability and the \acrshort{JRDCCP} metric, obtained with stochastic geometry and Monte-Carlo simulations, which are superimposed with each other. Following expressions in \eqref{eq:PD_PFA}, the false alarm probability is always lower than the detection probability at any threshold. The false alarm and detection probability decrease to zero when the radar threshold increases. Therefore, the number of false alarms is lowered at the price of less detections.\smallskip

For low values of $\threshRd$ (corresponding to high false alarm probabilities), the detection probability is close to one, and the \acrshort{JRDCCP} metric tends to be equal to the communication coverage probability. Contrariwise, for low values of the communication \acrshort{SIR} threshold $\threshC$, the communication coverage probability is close to one and the \acrshort{JRDCCP} metric tends to be equal to the detection probability. The same behaviour is obtained with the \acrshort{JRSCCP} metric, shown in Figure \ref{fig:JRSCCP}. Again, results obtained with stochastic geometry and Monte-Carlo simulations are superimposed. For low values of the radar \acrshort{SIR} threshold $\threshRs$, the radar success probability is close to one and the \acrshort{JRSCCP} tends to be equal to the coverage probability. By contrast, for low values of $\threshC$, the coverage probability is close to one, and the \acrshort{JRSCCP} tends to be equal to the radar success probability. Additionally, the condition given in \eqref{eq:JRSCCP_cond} is verified: with a radar processing gain set to one, whatever the values of $\threshRs$ and $\threshC$, the metric decreases to zero when the sum (in dB) exceeds 0dB. Owing to the negative impact of the communication signal on the radar function and vice-versa, this directly expresses that a trade-off should be achieved. \smallskip

Two conditional metrics are also illustrated in Figure \ref{fig:JRSCCP}d, in which the same condition applies: among the vehicle achieving at least a communication (resp. radar) \acrshort{SIR} $\threshC$ (resp. $\threshRs$) in dB, there is a zero probability of achieving a radar (resp. communication) \acrshort{SIR} higher than -$\threshC$ (resp. -$\threshRs$) in dB.\smallskip

Upper and lower bounds are illustrated in both graphs for one curve (dark blue). In both cases, the bounds become tighter as the joint metrics are evaluated for lower thresholds. In these cases, the joint metrics tend to be equal to radar or communication metrics (detection, success and coverage probabilities) and the bounds merge on the metrics.\smallskip

\subsection{Optimisation of the power levels}

\begin{figure}
\centering
\includegraphics[width=0.9\linewidth]{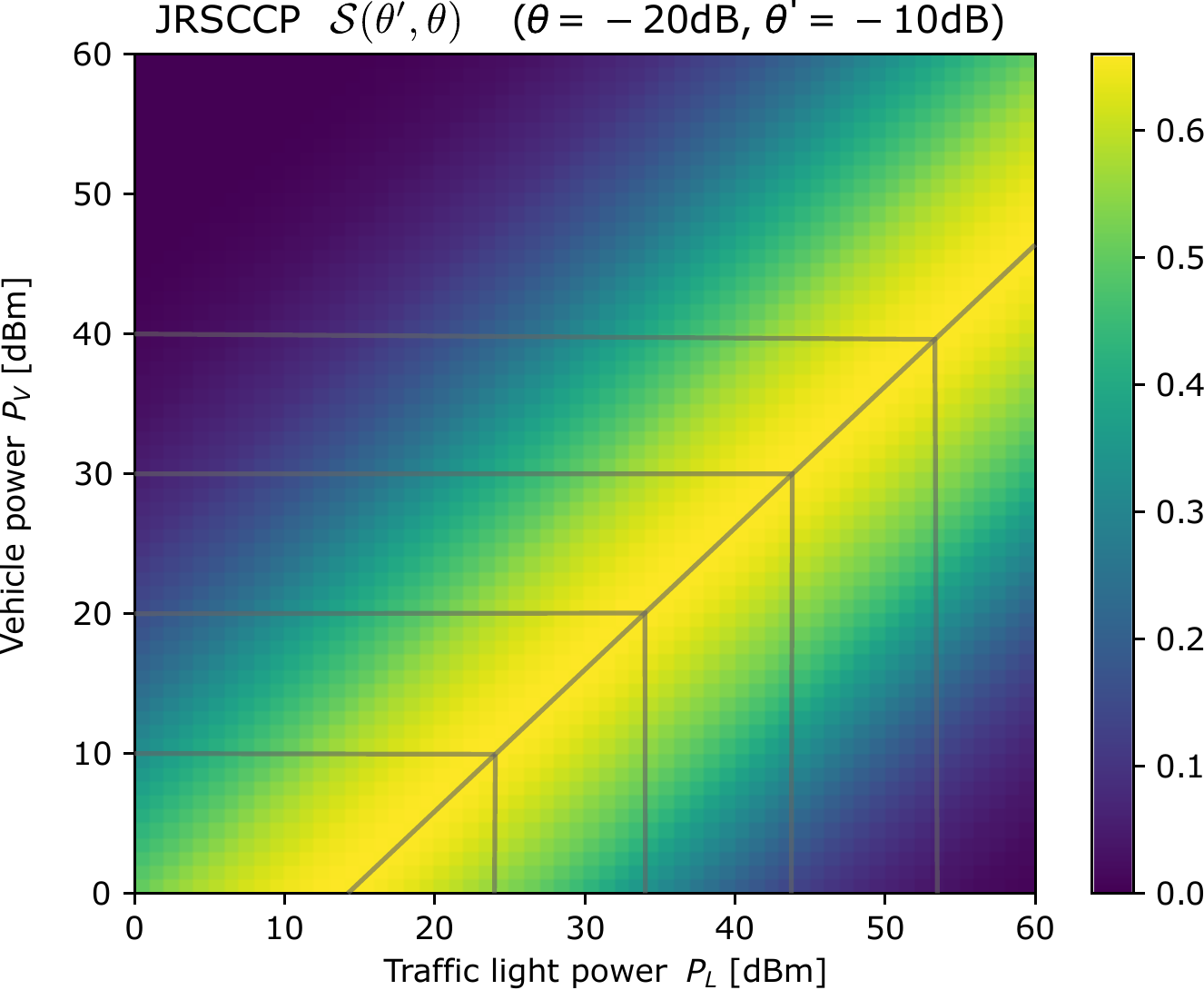}
\caption{\acrshort{JRSCCP} vs transmitted powers, with $\threshRs = -10$dB and $\threshC= -20$dB.}
\label{fig:JRSCCP_power}
\end{figure}

\begin{figure}
\centering
\includegraphics[width=0.9\linewidth]{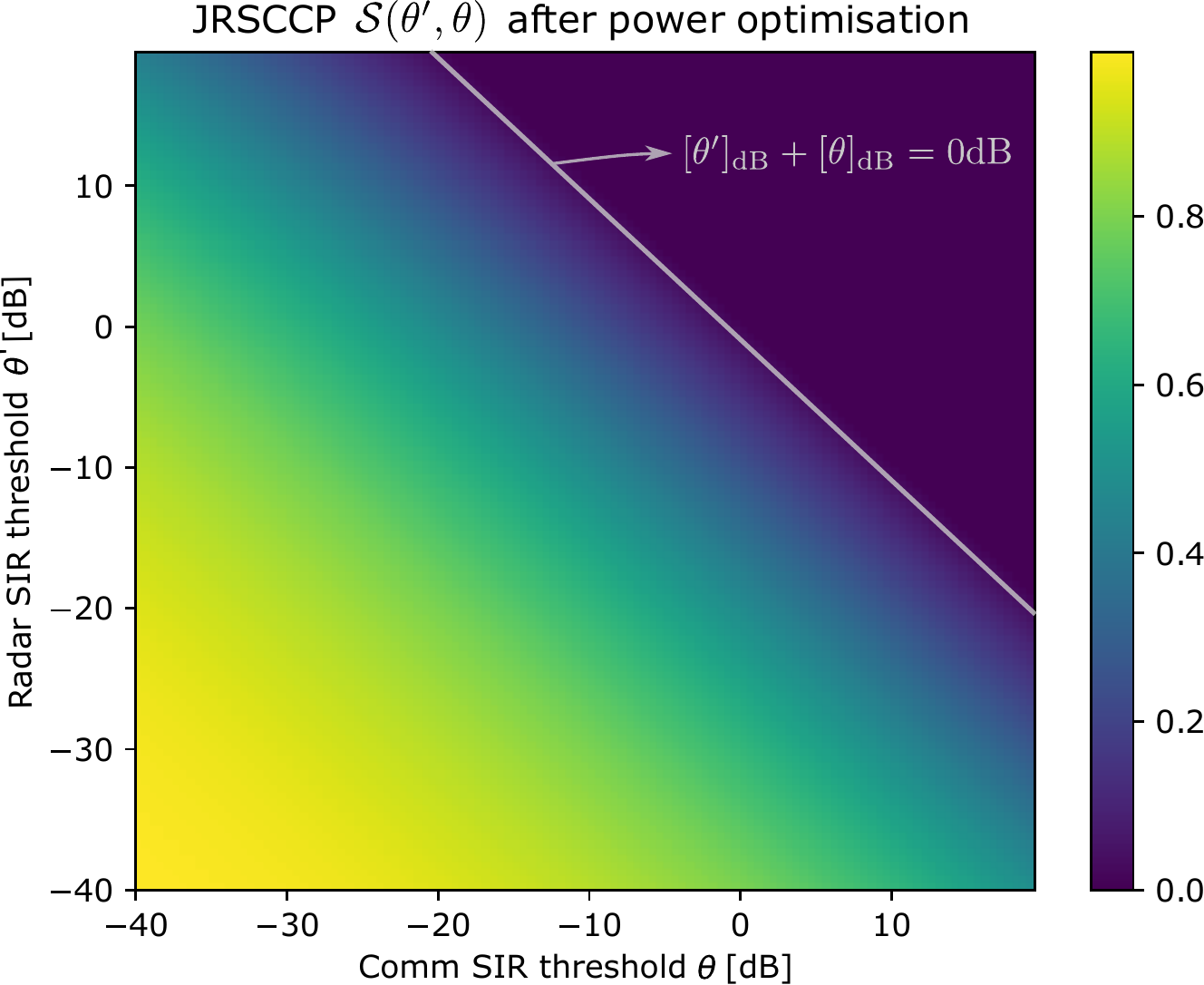}
\caption{\acrshort{JRSCCP} for different radar and communication \acrshort{SIR} thresholds $\threshRs$ and $\threshC$, with traffic light and vehicle powers optimised for each pair of thresholds. The minimum and maximum powers are $\Plm = \Pvm = 0$dBm and $\PlM = \PvM = 0$dBm.}
\label{fig:JRSCCP_power_max}
\end{figure}

Using the joint metrics, optimisation w.r.t. the traffic light and vehicle powers can be performed. For instance, with the \acrshort{JRSCCP} metric, the optimisation problem is formulated for given radar and communication \acrshort{SIR} thresholds $\threshRs$ and $\threshC$ as
\begin{equation}
\begin{alignedat}{10}
&\Pl^*, \Pv^* &&=\:&& \underset{\Pl,\Pv}{\arg\max} \quad &&\JRSCCP(\threshRs,\threshC ; \Pl, \Pv) \\
&&&&& \text{s.t.} && \Plm \leq \Pl \leq \PlM, \\
&&&&&             && \Pvm \leq \Pv \leq \PvM.
\end{alignedat}
\end{equation}
Minimum and maximum values for both traffic light and vehicle powers are respectively denoted by $\Plm$, $\PlM$, $\Pvm$ and $\PvM$.\smallskip

Figure \ref{fig:JRSCCP_power} illustrates how the \acrshort{JRSCCP} metric evolves for different transmitted powers $\Pv$ and $\Pl$ at given \acrshort{SIR} thresholds $\threshC$ and $\threshRs$. The power of the smart traffic lights should increase linearly in dB scale with the vehicle power to optimise the joint metric. Indeed, under the interference-limited scenario assumption, the \acrshort{JRSCCP} metric can be expressed as a function of the ratio between the traffic light and vehicle powers. \smallskip

For a given vehicle power, if the communication power is too low, the communication function does not perform well and the joint metric decreases. However, if the communication power is too high, the communication function performs well, but the radar function is impacted and the metric also decreases. Note that an increase of the vehicle power also leads to an increased interference for the radar function, while an increase of the traffic light power only improves the communication \acrshort{SIR} in the considered scenario. Note that similar observations can be performed with the traffic light density: an increased density leads to a higher probability for the nearest traffic light to be close. Therefore, the communication power is higher, increasing the communication \acrshort{SIR}, but also the interference at the radar function.\smallskip

The achieved JRSCCP metric after power optimisation is illustrated in Figure \ref{fig:JRSCCP_power_max}, for power levels comprised between 0 and 60dBm. Indeed, the achieved performance is higher when low \acrshort{SIR} thresholds are required. The condition of \eqref{eq:JRSCCP_cond} is also visible: the \acrshort{JRSCCP} metric cannot exceed zero if the \acrshort{SIR} threshold requirements are too high. Surprisingly, the metric is symmetric w.r.t. each threshold. It is observed when the interference is low compared to the radar and communication signals. This is happening in most cases with parameters of Table \ref{tab:scenar_params}.

\section{Extension with Interference Cancellation}
\label{sec:interference_cancellation}
From the radar point of view, any communication signal arriving at the receiver is an interfering signal. From the communication point of view, any unwanted communication signal reaching the receiver is interfering, but the radar echoes coming back to the DFRC transmitter are also considered as interference. Enabling some cooperation between both functions, communication signals which are decoded at the communication receiver could be (partially) suppressed at the radar receiver, and radar echoes which are processed at the radar receiver could be (partially) suppressed at the communication receiver. The higher the SINR at the communication receiver, the better the interference cancellation at the radar receiver. Similarly, the higher the SINR at the radar receiver, the better the interference cancellation at the communication receiver. 

In order to introduce interference cancellation, we consider the scenario where the communication signal arising from the closest smart traffic light is the only communication signal which is decoded. Therefore, the interference arising from the opposite lane is not cancelled, but the cooperation of both functions enables to mitigate the impact of $\Sr$ on the communication function, and $\Sc$ on the radar function. In that case, denoting by $\resR(\Sr,\Sc) \equiv \resR$ and $\resC(\Sr,\Sc) \equiv \resC$ the residual interference after interference cancellation, the metrics presented in Sections \ref{sec:PC} to \ref{sec:JRSCCP} are respectively rewritten as 
\begin{align}
&\PCPIC(\threshC) = \Pr{}{\frac{\Sc}{\resR + \I} \geq \threshC},\\
&\PFAPIC(\threshRd) = \Pr{\Sc,\I}{\resC+\I\geq \threshRd},\\
&\PDPIC(\threshRd) = \Pr{}{\Sr\RPG + \resC + \I \geq \threshRd}, \\
&\PSPIC(\threshRs) = \Pr{}{\frac{\Sr \RPG}{\resC + \I} \geq \threshRs},\\
    &\JRDCCPPIC(\threshC;\threshRd) = \Pr{}{\Sr\RPG + \resC + \I \geq \threshRd, \: \frac{\Sc}{\resR + \I} \geq \threshC}  \\
\nonumber &\qquad\qquad\text{for a given } \PFAPIC(\threshRd), \\
    &\JRSCCPPIC(\threshRs,\threshC) = \Pr{}{\frac{\Sr\RPG}{\resC + \I} \geq \threshRs,\: \frac{\Sc}{\resR + \I} \geq \threshC}.
\end{align}
The residual interference $\resR$ is supposed to be high when the radar SIR is low since the radar receiver has difficulties to estimate correctly the delay and Doppler frequencies of the different targets. When this SIR is close to 0dB, the interference level could even be increased. Similarly, the same applies for $\resC$ when the communication SIR is low since the communication receiver has difficulties to decode the communication signal and to estimate the associated parameters. Conversely, when these ratios are high, the interference is efficiently cancelled, leading to nearly zero residuals. For tractability, even if the total interference should be considered to assess the performance achieved by the interference cancellation, we assume that these functions do not depend on the aggregate interference $\I$.

In this paper, three different cases are evaluated.
\begin{enumerate}
    \item No interference cancellation is performed:
    \begin{equation*}
        \resR(\Sr,\Sc) = \Sr, \quad \resC(\Sr,\Sc) = \Sc,
    \end{equation*}
    leading back to definitions given in Section \ref{sec:metrics_development}.
    \item An imperfect interference cancellation is performed at both functions simultaneously, leading to the following residuals:
    \begin{align*}
        \resR(\Sr,\Sc) &= \Sr\:\res\left(\frac{\Sr}{\Sc}\right),\\
        \resC(\Sr,\Sc) &= \Sc\:\res\left(\frac{\Sc}{\Sr}\right),
    \end{align*}
    with 
    \begin{equation*}
     \res(x) = \frac{a}{1+x^b}.   
    \end{equation*}
    This function is illustrated in Figure \ref{fig:zeta_func} for different values of $a$ and $b$. It has been designed to integrate the positive and negative impacts of interference cancellation depending on the power levels $\Sr$ and $\Sc$. Choosing $a > 1$ takes into account an increase of the interference for low power ratios.
    
    \begin{figure}
        \centering
        \includegraphics[width=0.95\linewidth]{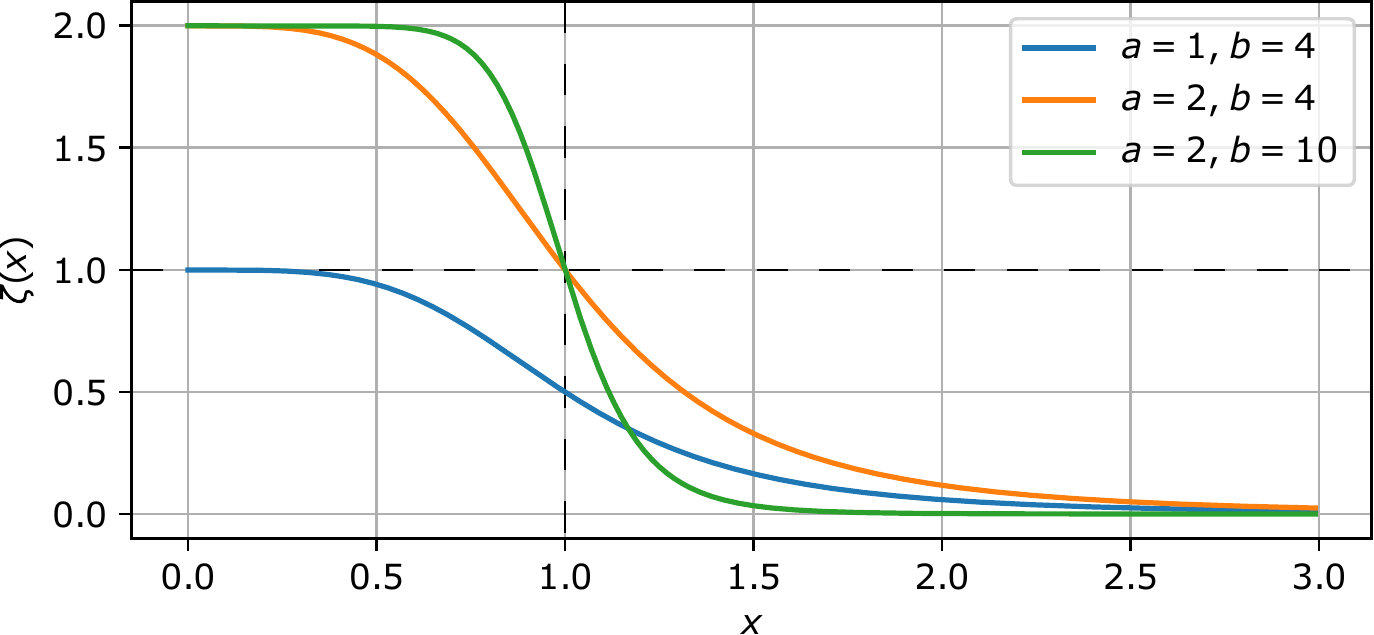}
        \caption{Function $\res$ for different parameters.}
        \label{fig:zeta_func}
    \end{figure}
    
    \item A perfect interference cancellation is performed at both functions simultaneously, leading to null residuals:
    \begin{equation*}
        \resR(\Sr,\Sc) = 0 \quad\text{and/or}\quad \resC(\Sr,\Sc) = 0
    \end{equation*}
    depending on whether the interference cancellation is performed at the radar receiver, at the communication receiver or at both functions.
\end{enumerate}

Propositions \ref{prop:PC_IC} to \ref{prop:JRSCCP_IC} extend the results of Section \ref{sec:metrics_development} to integrate interference cancellation in the metrics.

\proposition{PC_IC}{For a given communication SIR threshold $\threshC$, the coverage probability with interference cancellation is obtained as
{\normalfont
\begin{equation}
\PCPIC(\threshC) = \restesp{\setPCPIC}{\frac{1}{2}-\frac{1}{\pi}\int_0^\infty \frac{1}{\tau}\:\Im{\LT_\I(-j\tau)\:\funcPCPIC\left(\tau\right)}\d \tau}
\label{eq:CP_eq_IC}
\end{equation}
}
where
{\normalfont
\begin{equation}
\funcPCPIC(\tau) = \exp{-j\tau \left(\frac{\Sc}{\threshC} - \resR\right)}
\label{eq:CP_func_IC}
\end{equation}
}
and the area $\setPCPIC$ is defined as
{\normalfont
\begin{equation}
\setPCPIC \triangleq \left\{
\begin{alignedat}{10}
&\Sc - \theta \resR \geq 0\\
&\Srmin \leq \Sr \leq \Srmax \\
&0 \leq \Sc \leq \Scmax
\end{alignedat}
\right\}.
\label{eq:CP_set_IC}
\end{equation}
}
}{The proof is detailed in Appendix \ref{app:proof_CP}.}

\proposition{PFAPIC}{For a given threshold $\threshRd$, the false alarm probability with interference cancellation is given by
{\normalfont
\begin{align}
    \PFAPIC(\threshRd) &= 1 - \restesp{\setPFAPIC}{\frac{1}{2} - \frac{1}{\pi}\int_0^\infty \frac{1}{\tau}\:\Im{\LT_\I(-j\tau)\:\funcPFAPIC(\tau)}\d \tau} 
    \label{eq:FAP_eq_IC}
\end{align}
}
where
{\normalfont
\begin{equation}
    \funcPFAPIC(\tau) = \exp{-j\tau(\threshRd - \resC)},
    \label{eq:FAP_func_IC}
\end{equation}
}
and the area $\setPFAPIC$ is defined as
{\normalfont
\begin{equation}
\setPFAPIC \triangleq \left\{
\begin{alignedat}{10}
&\resC \leq \threshRd \\
&\Srmin \leq \Sr \leq \Srmax \\
&0 \leq \Sc \leq \Scmax
\end{alignedat}
\right\}.
\label{eq:FAP_set_IC}
\end{equation}
}
}{The proof is detailed in Appendix \ref{app:proof_FAP}.}

\proposition{PDPIC}{For a given threshold $\threshRd$, the detection probability with interference cancellation is given by
{\normalfont
\begin{align}
    &\PDPIC(\threshRd) = 1 - \restesp{\setPDPIC}{\frac{1}{2} - \frac{1}{\pi}\int_0^\infty \frac{1}{\tau}\:\Im{\LT_\I(-j\tau)\:\funcPDPIC(\tau)}\d\tau}
    \label{eq:DP_eq_IC}
\end{align} 
}
where
{\normalfont
\begin{equation}
    \funcPDPIC(\tau) = \exp{-j\tau(\threshRd - \Sr\RPG - \resC)},
    \label{eq:DP_func_IC}
\end{equation}
}
and the area $\setPDPIC$ is defined as
{\normalfont
\begin{equation}
\setPDPIC \triangleq \left\{
\begin{alignedat}{10}
&\threshRd - \Sr\RPG - \resC \geq 0\\
&\Srmin \leq \Sr \leq \Srmax \\
&0 \leq \Sc \leq \Scmax
\end{alignedat}
\right\}.
\label{eq:DP_set_IC}
\end{equation}
}
}{The proof is detailed in Appendix \ref{app:proof_DP}.}

\proposition{PSPIC}{For a given radar \acrshort{SIR} threshold $\threshRs$, the success probability with interference cancellation is obtained as
{\normalfont
\begin{align}
&\PSPIC(\threshRs) = \restesp{\setPSPIC}{\frac{1}{2} - \frac{1}{\pi}\int_0^\infty \frac{1}{\tau}\:\Im{\LT_\I(-j\tau)\:\funcPSPIC(\tau)}\:\mathrm{d}\tau}
\label{eq:SP_eq_IC}
\end{align}
}
where
{\normalfont
\begin{equation}
\funcPSPIC(\tau) = \exp{-j\tau\left(\frac{\Sr\RPG}{\threshRs} - \resC\right)},
\label{eq:SP_func_IC}
\end{equation}
}
and the area $\setPSPIC$ is defined as
{\normalfont
\begin{equation}
\setPSPIC \triangleq \left\{
\begin{alignedat}{10}
&\resC \leq \Sr\RPG/\threshRs \\
&\Srmin \leq \Sr \leq \Srmax \\
&0 \leq \Sc \leq \Scmax
\end{alignedat}
\right\}.
\label{eq:SP_set_IC}
\end{equation}
}
}{The proof is detailed in Appendix \ref{app:proof_SP}.}

\proposition{JRDCCP_IC}{For a given false alarm probability $\PFA$ (and therefore for a given threshold $\threshRd$) and a given communication \acrshort{SIR} threshold $\threshC$, the \acrshort{JRDCCP} metric with interference cancellation is obtained as
{\normalfont
\begin{align}
    \nonumber & \JRDCCPPIC(\threshC;\threshRd) =  - \restesp{\setJRDCCPBPIC}{\frac{1}{\pi}\int_0^\infty  \frac{1}{\tau}\:\Im{\LT_\I(-j\tau)\:\funcJRDCCPBPIC(\tau)}\d\tau} \\
     & + \restesp{\setJRDCCPAPIC}{\frac{1}{2} -\frac{1}{\pi}\int_0^\infty\frac{1}{\tau}\:\Im{\LT_\I(-j\tau)\: \funcJRDCCPAPIC(\tau)}\d\tau} \label{eq:JRDCCP_eq_IC},
\end{align}
}
where
{\normalfont
\begin{equation}
    \funcJRDCCPAPIC(\tau) = \funcPCPIC(\tau),\quad \funcJRDCCPBPIC(\tau) = \funcPCPIC(\tau) - \funcPDPIC(\tau),
    \label{eq:JRDCCP_func_IC}
\end{equation}
}
and the areas $\setJRDCCPAPIC$ and $\setJRDCCPBPIC$ are defined as
{\normalfont
\begin{align}
    \setJRDCCPAPIC &\triangleq \left\{
    \begin{alignedat}{10}
    &\threshRd - \Sr\RPG - \resC \leq 0 \\
    &\frac{\Sc}{\threshC} - \resR \geq 0\\ 
    &\Srmin \leq \Sr \leq \Srmax \\
    &0 \leq \Sc \leq \Scmax
    \end{alignedat}
    \right\}, \\
    \setJRDCCPBPIC &\triangleq \left\{
    \begin{alignedat}{10}
    &\threshRd - \Sr\RPG - \resC \geq 0 \\
    &\frac{\Sc}{\threshC} - \resR \geq \threshRd - \Sr\RPG - \resC \\
    &\Srmin \leq \Sr \leq \Srmax \\
    &0 \leq \Sc \leq \Scmax
    \end{alignedat}
    \right\}.
    \label{eq:JRDCCP_set_IC}
\end{align}
}}{
The proof is detailed in Appendix \ref{app:proof_JRDCCP}.
}

\proposition{JRSCCP_IC}{For a given radar \acrshort{SIR} threshold $\threshRs$ and communication \acrshort{SIR} threshold $\threshC$, the \acrshort{JRSCCP} metric with interference cancellation is obtained as
{\normalfont
\begin{align}
    \nonumber &\JRSCCPPIC(\threshRs,\threshC) = \restesp{\setJRSCCPAPIC}{\frac{1}{2} -\frac{1}{\pi} \int_0^\infty \frac{1}{\tau}\:\Im{\LT_\I(-j\tau) \:\funcJRSCCPAPIC(\tau)} \d \tau} \\
    +& \restesp{\setJRSCCPBPIC}{\frac{1}{2} -\frac{1}{\pi} \int_0^\infty \frac{1}{\tau}\:\Im{\LT_\I(-j\tau) \:\funcJRSCCPBPIC(\tau)} \d \tau}\label{eq:JRSCCP_eq_IC},
\end{align}
}
where 
{\normalfont
\begin{equation}
    \funcJRSCCPAPIC(\tau) = \funcPSPIC(\tau), \quad \funcJRSCCPBPIC(\tau) = \funcPCPIC(\tau),
   \label{eq:JRSCCP_func_IC}
\end{equation}
}
and the areas $\setJRSCCPAPIC$ and $\setJRSCCPBPIC$ are defined as
{\normalfont
\begin{align}
    \setJRSCCPAPIC &\triangleq \left\{
    \begin{alignedat}{10}
    &\frac{\Sr\RPG}{\threshRs} - \resC \geq 0 \\
    &\frac{\Sc}{\threshC} - \resR \geq \frac{\Sr\RPG}{\threshRs} - \resC\\
    &\Srmin \leq \Sr \leq \Srmax \\
    &0 \leq \Sc \leq \Scmax
    \end{alignedat}
    \right\}, \\
    \setJRSCCPBPIC &\triangleq \left\{
    \begin{alignedat}{10}
    & \frac{\Sc}{\threshC} - \resR \geq 0 \\
    & \frac{\Sr\RPG}{\threshRs} - \resC \geq \frac{\Sc}{\threshC} - \resR\\
    &\Srmin \leq \Sr \leq \Srmax \\
    &0 \leq \Sc \leq \Scmax
    \end{alignedat}
    \right\}.
    \label{eq:JRSCCP_set_IC}
\end{align}
}
}{The proof is detailed in Appendix \ref{app:proof_JRSCCP}.}

\begin{figure*}
	\centering
	\includegraphics[width=1.0\linewidth]{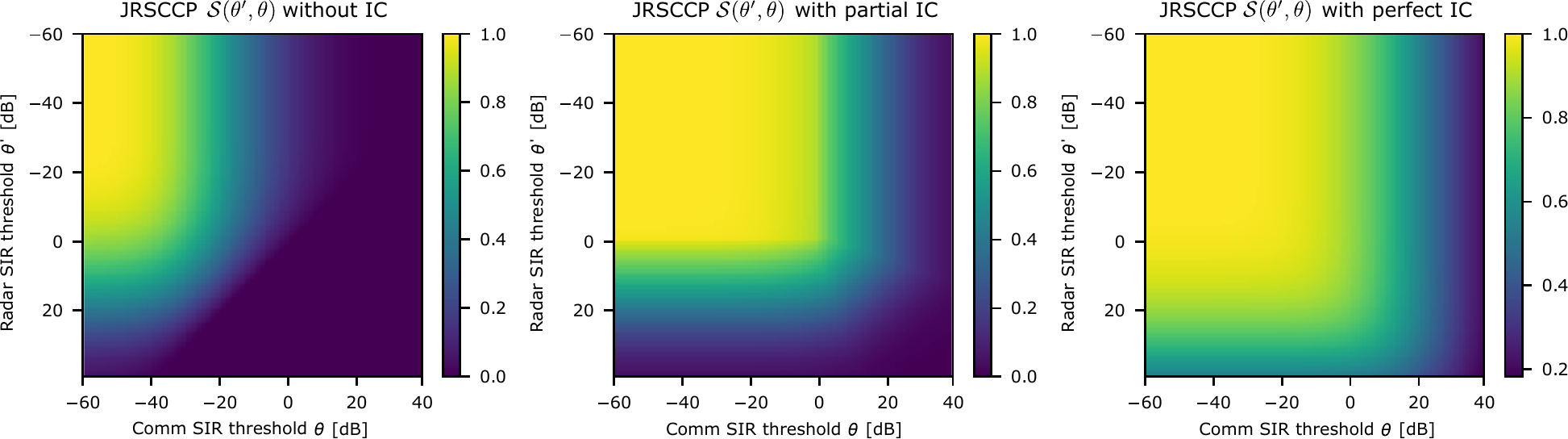}
	\caption{JRSCCP metric with different interference cancellation schemes.}
    \label{fig:JRSCCP_IC_2D}
\end{figure*}

\begin{figure}
    \centering
    \includegraphics[width=0.9\linewidth]{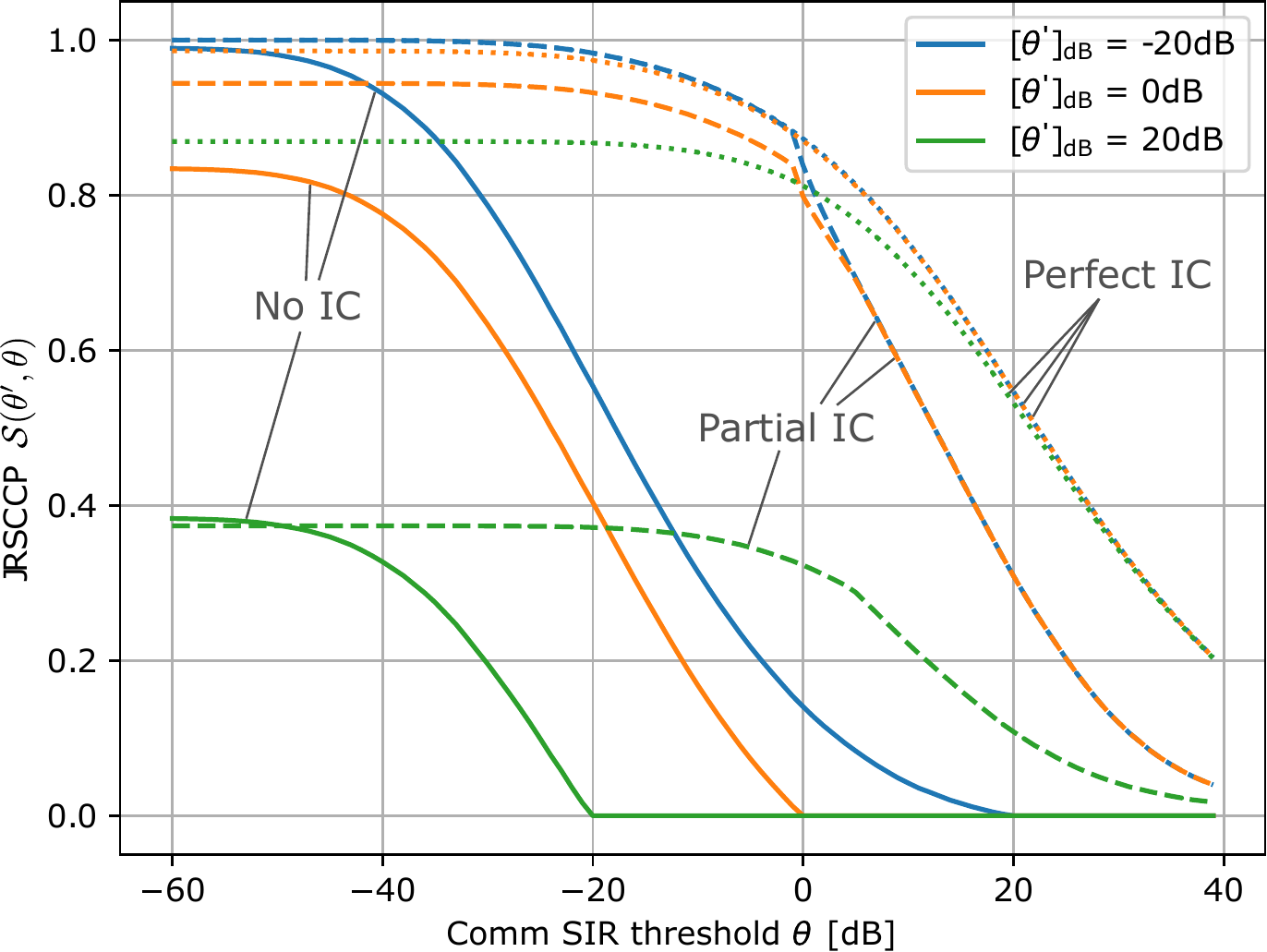}
    \caption{JRSCCP metric with different interference cancellation schemes, for $\threshRs = \{-20,0,20\}\text{dB}$.}
    \label{fig:JRSCCP_IC}
\end{figure}

The upper and lower bounds are also generalised for the joint metrics with interference cancellation in Corrolaries \ref{corr:JRDSCCPIC_bounds} using Fréchet inequalities.

\corrolary{JRDCCPIC_bounds}
{
For a given false alarm probability $\PFAPIC$ with interference cancellation (and therefore for a given threshold $\threshRd$) and a given communication SIR threshold $\threshC$, the JRDCCP metric with interference cancellation is bounded by
{\normalfont
\begin{equation}
\JRDCCPPICmin(\threshC;\threshRd) \leq \JRDCCPPIC(\threshC;\threshRd) \leq \JRDCCPPICmax(\threshC;\threshRd),
\end{equation}
}
with
{\normalfont
\begin{align}
\JRDCCPPICmin(\threshC;\threshRd) &= \max\left(0,\PDPIC(\gamma) + \PCPIC(\threshC) -1 \right), \\
\JRDCCPPICmax(\threshC;\threshRd) &= \min\left(\PDPIC(\threshRd),\PCPIC(\threshC)\right).
\end{align}
}
Similarly, for a given radar \acrshort{SIR} threshold $\threshRs$ and communication \acrshort{SIR} threshold $\threshC$, the JRSCCP metric is bounded by
{\normalfont
\begin{equation}
\JRSCCPPICmin(\threshRs,\threshC) \leq \JRSCCPPIC(\threshRs,\threshC) \leq \JRSCCPPICmax(\threshRs,\threshC),
\end{equation}
}
with
{\normalfont
\begin{align}
\JRSCCPPICmin(\threshRs,\threshC) &= \max\left(0,\PSPIC(\threshRs) + \PCPIC(\threshC) -1 \right), \\
\JRSCCPPICmax(\threshRs,\threshC) &= \min\left(\PSPIC(\threshRs),\PCPIC(\threshC)\right).
\end{align}
}
}
{
The proposed bounds are given by Fréchet inequalities.
}

Finally, the conditional metrics are rewritten as follows:
\begin{align}
\PCifDPIC(\threshC;\threshRd) &= \Pr{}{\frac{\Sc}{\resR + \I} \geq \threshC \:\bigg|\: \Sr\RPG + \resC + \I \geq \threshRd} \\
    \nonumber &\text{for a given } \PFAPIC(\threshRd), \\
\PCifSPIC(\threshC;\threshRs) &= \Pr{}{\frac{\Sc}{\resR + \I} \geq \threshC \:\bigg|\: \frac{\Sr\RPG}{\resC + \I} \geq \threshRs }. \\
\PDifCPIC(\threshRd;\threshC) &= \Pr{}{\Sr\RPG + \resC + \I \geq \threshRd \:\bigg|\: \frac{\Sc}{\resR + \I} \geq \threshC} \\
    \nonumber &\text{for a given } \PFAPIC(\threshRd), \\
\PSifCPIC(\threshRs;\threshC) &= \Pr{}{\frac{\Sr\RPG}{\resC + \I} \geq \threshRs  \:\bigg|\: \frac{\Sc}{\resR + \I} \geq \threshC}.
\end{align}
Their expressions are given in Corrolaries \ref{corr:condMetCDSIC}.

\corrolary{condMetCDSIC}{For a given false alarm probability $\PFA$ (and therefore for a given threshold $\threshRd$) or a given radar \acrshort{SIR} threshold $\threshRs$, the coverage probabilities achieved by vehicles performing successfully the radar function with interference cancellation for a given communication \acrshort{SIR} threshold $\threshC$ are given by
{\normalfont
\begin{equation}
\PCifDPIC(\threshC;\threshRd) = \JRDCCPPIC(\threshC;\threshRd) / \PDPIC(\threshRd)
\end{equation}
}
and
{\normalfont
\begin{equation}
\PCifSPIC(\threshC;\threshRs) = \JRSCCPPIC(\threshRs,\threshC) / \PSPIC(\threshRs).
\end{equation}
}
Similarly, the detection and success probabilities achieved by vehicles performing successfully the communication function with interference cancellation are given by
{\normalfont
\begin{equation}
\PDifCPIC(\threshRd;\threshC) = \JRDCCPPIC(\threshC;\threshRd) / \PCPIC(\threshC)
\end{equation}
}
and
{\normalfont
\begin{equation}
\PSifCPIC(\threshRs;\threshC) = \JRSCCPPIC(\threshRs,\threshC) / \PCPIC(\threshC).
\end{equation}
}
}{
These results are obtained using Bayes law.
}

Figure \ref{fig:JRSCCP_IC_2D} and \ref{fig:JRSCCP_IC} illustrates the impact of the different interference cancellation schemes on the \acrshort{JRSCCP} metric: perfect interference cancellation, partial interference cancellation with $a=2$ and $b=4$, or no interference cancellation at all. Note that with this choice of parameters, the residual interference $\resR$ (resp. $\resC$) at the communication function (resp. radar function) is increased when the power ratio $\Sr/\Sc$ (resp. $\Sc/\Sr$) is less than 0dB, as illustrated in Figure \ref{fig:zeta_func}.\smallskip

With perfect interference cancellation, the performance drastically increases since the interference power is in most cases lower than the radar and communication powers.  With partial interference cancellation, the achieved performance depends on the considered radar and communication \acrshort{SIR} threshold. At high radar (resp. communication) \acrshort{SIR} thresholds, it is likely that the communication (resp. radar) signal is dominant for the metric to be non zero. In that situation, the interference cancellation performs well at the communication (resp. radar) function, but the interference level is potentially increased at the other function. However, in most cases, the achieved performance is improved thanks to partial interference cancellation.

\section{Conclusion}
In this paper, a new automotive scenario with smart traffic lights and vehicles, both equipped with joint radar-communication systems, has been analysed. In order to analyse the performance of both functions separately or jointly, multiple metrics, upper and lower bounds have been developed using stochastic geometry. Based on that, optimisation of the powers of the vehicles and traffic lights has also been performed. Finally, the metrics have been extended to include a new model of interference cancellation at both functions.\smallskip

In future works, other models and schemes for the interference cancellation could be considered, i.e. suppressing the interference at the radar or communication function only. Additionally, both metrics could be generalised for other joint radar-communication scenarios, for instance indoor communication and localisation. Finally, this work can be extended by comparing the developed metrics with these obtained through Monte-Carlo simulations with a complexified system model, and ray tracing simulations.

\bibliographystyle{ieeetr}
\bibliography{biblio}

\appendices

\section{Probability density functions of $\rR$ and $\rC$}
\label{app:proof_PDF_r}
Let us denote by $\RCM{\PhiR}(A)$ the random counting measure of $\PhiR$ on the subset $A$. Knowing that the void probability of a \acrshort{PPP} of density $\lambda$ is given by
\begin{equation*}
    \Pr{\RCM{\Phi}}{\RCM{\Phi}(A) = 0} = \exp{-\int_A \lambda(\textbf{x})\d\textbf{x}},
\end{equation*}
The probability to have at least one vehicle in the radar detectable range is given by
\begin{multline*}
    \Pr{\RCM{\PhiR}}{\RCM{\PhiR}([\rRmin,\rRmax]) > 0}
    \\ = 1 - \exp{-\lambdaV(\rRmax - \rRmin)}.
\end{multline*}
Knowing that, thanks to the Bayes rule, the complementary \acrshort{CDF} of $\rR$ assuming that there is at least one vehicle in the radar detection range can be developed as
\begin{align*}
    &\CCDF{\rR}{r} = \Pr{\rR,\RCM{\PhiR}}{\rR > r\:|\:\RCM{\PhiR}([\rRmin,\rRmax]) > 0}\\
    &= \frac{\Pr{\rR,\RCM{\PhiR}}{\RCM{\PhiR}([0,r]) = 0, \RCM{\PhiR}([\rRmin,\rRmax]) > 0}}{\Pr{\RCM{\PhiR}}{\RCM{\PhiR}([\rRmin,\rRmax]) > 0}}.
\end{align*}
Three cases can be distinguished:
\begin{enumerate}
    \item When $r \leq \rRmin$, the two subsets $[0,r]$ and $[\rRmin,\rRmax]$ are disjoint, and therefore, from the properties of \acrshort{PPP}s, the counting measures evaluated on these two subsets are independent. Thus,
    \begin{equation*}
        \CCDF{\rR}{r} = \Pr{\rR,\RCM{\PhiR}}{\RCM{\PhiR}([0,r]) = 0} = 1.
    \end{equation*}
    \item When $\rRmax \leq r$, $[\rRmin,\rRmax]$ is a subset of $[0,r]$, and therefore $\CCDF{\rR}{r} = 0$.
    \item When $\rRmin \leq r \leq \rRmax$, based on the void probability of \acrshort{PPP}s, the numerator of the above expression can be developed as 
    \begin{align*}
        &\Pr{\rR,\RCM{\PhiR}}{\RCM{\PhiR}([0,r]) = 0, \RCM{\PhiR}([\rRmin,\rRmax]) > 0} \\
        &= \Pr{\rR,\RCM{\PhiR}}{\RCM{\PhiR}([\rRmin,r]) = 0, \RCM{\PhiR}([r,\rRmax]) > 0}  \\
        &= \exp{-\lambdaV(r-\rRmin)}[1-\exp{-\lambdaV(\rRmax - r}].
    \end{align*}
\end{enumerate}
Finally the \acrshort{PDF} of $\rR$ is obtained by derivating the opposite of the cumulative \acrshort{CDF}, and \ref{eq:PDFrR} is obtained. Equation \eqref{eq:PDFrC} can be demonstrated similarly.

\section{Probability density functions of $\Sr$ and $\Sc$}
\label{app:proof_PDF_s}
These results directly follow from the change of variable of an univariate \acrshort{PDF} with a monotonic and invertible function: let $g : A\in\mathbb{R} \rightarrow B\in\mathbb{R}$ be a monotonic and invertible function and $Y = g(X)$ with $X$ and $Y$ two random variables, the \acrshort{PDF} of $Y$ is obtained as
\begin{equation*}
    \PDF{Y}{y} = \PDF{X}{g^{-1}(y)}\:\left|\frac{\d}{\d y}g^{-1}(y)\right|,
\end{equation*}
where $\PDF{X}{x}$ is the \acrshort{PDF} of $X$ and $g^{-1}$ denotes the inverse function of $g$. For the radar link, \eqref{eq:SrRT} is inverted as $\rR(s) = \rhoR^{\frac{1}{\PLE_R}}\:s^{-\frac{1}{\PLE_R}}$, and 
\begin{equation*}
    \left|\frac{\d}{\d s}\rR(s)\right| = \frac{\rhoR^{\frac{1}{\PLE_R}} \: s^{-\frac{1}{\PLE_R}-1}}{\PLE_R} = \frac{\rR(s)}{\PLE_R\:s},
\end{equation*}
leading to \eqref{eq:PDFsRrt}. For the communication link, \eqref{eq:Sc} is inverted as $\rC(s) = \sqrt{\rhoC^{\frac{2}{\PLE_C}}\:s^{-\frac{2}{\PLE_C }}-\dC^2}$,
and 
\begin{align*}
    \left|\frac{\d}{\d s}\rC(s)\right| &= \frac{2\rhoC^{\frac{2}{\PLE_C}}\:s^{-\frac{2}{\PLE_C}-1}}{\PLE_C}\frac{1}{2\sqrt{\rhoC^{\frac{2}{\PLE_C}}\:s^{-\frac{2}{\PLE_C}}-\dC^2}} \\
    &= \frac{\rC^2(s) + \dC^2}{\PLE_C\:s\:\rC(s)},
\end{align*}
leading to \eqref{eq:PDFsC}.

\section{Laplace transform of $|h_i|^2$}
\label{app:proof_LT_h}
Let us denote by $\nu^2$ the power of the dominant path and $2\sigma^2$ the power of the scattered paths of an interfering link. The small-scale fading is expressed in cartesian coordinates as $|h_i|^2 = X^2+Y^2$ where $X\sim\mathcal{N}(\nu\cos\theta,\sigma^2)$ and $Y\sim\mathcal{N}(\nu\sin\theta,\sigma^2)$ are independent normally distributed random variables, with $\theta\in [0,2\pi[$. By the properties of the normal distribution, it is known that $X^2/\sigma^2$ and $Y^2/\sigma^2$ follow chi-square distributions with one degree of freedom. Therefore we have
\begin{equation*}
    \LT_{X^2}(s) = \exp{-\frac{\nu^2\cos^2\theta}{1+2\sigma^2\:s}\:s} \left(1+2\sigma^2\:s\right)^{-\frac{1}{2}}
\end{equation*}
and
\begin{equation*}
    \LT_{Y^2}(s) = \exp{-\frac{\nu^2\sin^2\theta}{1+2\sigma^2\:s}\:s} \left(1+2\sigma^2\:s\right)^{-\frac{1}{2}}.
\end{equation*}
Finally, \eqref{eq:LTh} is obtained knowing that $\LT_{|h_i|^2}(s) = \LT_{X^2}(s)\LT_{Y^2}(s)$, 
\begin{equation*}
    \nu^2 = \frac{K}{K+1}\:\Omega,\quad 2\sigma^2 = \frac{1}{K+1}\:\Omega,
\end{equation*}
and $\esp{}{|h_i|^2} = \Omega = 1$ with a normalised power.

\section{Laplace transform of $\I$}
\label{app:proof_LT_I}
By developing the definition of the Laplace transform of $\I$,
\begin{align*}
&\LT_\I(s) = \esp{}{\exp{-s\sum_{i\:|\:\phi_i \in \PhiI} \rhoI \left(\rVert\phi_i\rVert^2 + \dI^2\right)^{-\frac{\PLE_I}{2}} |h_i|^2}} \\
&\overset{(a)}{=} \esp{\PhiI}{\prod_{i\:|\:\phi_i \in \PhiI} \esp{h_i}{\exp{-s \rhoI \left(\rVert\phi_i\rVert^2 + \dI^2\right)^{-\frac{\PLE_I}{2}} |h_i|^2}}} \\
&\overset{(b)}{=} \exp{-\lambdaI\int_{\rImin}^\infty \left(1-\LT_{|h_i|^2}\left(s \rhoI \left(r^2 + \dI^2\right)^{-\frac{\PLE_I}{2}}\right)\right)\d r}
\end{align*}
where $(a)$ is obtained thanks to the independency of the small-scale fading of each interfering link, and $(b)$ follows from the Probability Generating Functional of a \acrshort{PPP}.

\section{JRDCCP and derived metrics}
\subsection{JRDCCP metric}
\label{app:proof_JRDCCP}
\subsubsection{With interference cancellation}
The \acrshort{JRDCCP} metric can be developed as
\begin{align*}
    &\JRDCCPPIC(\threshC;\gamma) = \Pr{\Sr,\Sc,\I}{\Sr\RPG + \resC + \I \geq \threshRd, \frac{\Sc}{\resR + \I} \geq \threshC} \\
    &= \esp{\Sr,\Sc}{\Pr{\I|\Sr,\Sc}{\threshRd - \Sr\RPG - \resC \leq \I \leq \frac{\Sc}{\threshC} - \resR}}.
\end{align*}
This probability is non zero either when
\begin{equation}
    \left\{
    \begin{alignedat}{10}
    &\threshRd - \Sr\RPG - \resC \leq 0 \\
    &\frac{\Sc}{\threshC} - \resR \geq 0
    \end{alignedat}
    \right. \label{app:condJRDCCPA}
\end{equation}
or when 
\begin{equation}
    \left\{
    \begin{alignedat}{10}
    &\threshRd - \Sr\RPG - \resC \geq 0 \\
    &\frac{\Sc}{\threshC} - \resR \geq \threshRd - \Sr\RPG - \resC
    \end{alignedat}
    \right., \label{app:condJRDCCPB}
\end{equation}
leading to the areas $\setJRDCCPAPIC$ and $\setJRDCCPBPIC$ given in \eqref{eq:JRDCCP_set_IC} including the radar and communication power constraints.\smallskip

In $\setJRDCCPAPIC$, applying the Gil-Pelaez theorem,
\begin{multline*}
    \Pr{\I|\Sr,\Sc}{\I \leq \frac{\Sc}{\threshC} - \resR} = \\
      \frac{1}{2} - \frac{1}{\pi}\int_0^\infty  \frac{1}{\tau} \: \Im{\LT_\I(-j\tau) \: \funcJRDCCPAPIC(\tau)}\d\tau
\end{multline*}
with $\funcJRDCCPAPIC$ given in \eqref{eq:JRDCCP_func_IC}. 

In $\setJRDCCPBPIC$, applying the Gil-Pelaez theorem,
\begin{align*}
   &\Pr{\I|\Sr,\Sc}{\threshRd - \Sr\RPG - \resC \leq \I \leq \frac{\Sc}{\threshC} - \resR} \\
    &= \Pr{\I|\Sr,\Sc}{\I \leq \frac{\Sc}{\threshC} - \resR}\\
    &\qquad - \Pr{\I|\Sr,\Sc}{I \leq \threshRd - \Sr\RPG - \resC} \\
    &= -\frac{1}{\pi}\int_0^\infty  \frac{1}{\tau}\:\Im{\LT_\I(-j\tau)\:\funcJRDCCPBPIC(\tau)}\d\tau
\end{align*}
with $\funcJRDCCPBPIC$ given in \eqref{eq:JRDCCP_func_IC}, leading finally to \eqref{eq:JRDCCP_eq_IC}.

\subsubsection{Without interference cancellation}
Without interference cancellation, \eqref{app:condJRDCCPA} and \eqref{app:condJRDCCPB} are respectively rewritten as 
\begin{equation*}
    \left\{
    \begin{alignedat}{10}
    &\threshRd - \Sr\RPG - \Sc \leq 0 \\
    &\frac{\Sc}{\threshC} - \Sr \geq 0
    \end{alignedat}
    \right. \quad\Leftrightarrow\quad \Sc \geq \max\left(\threshC\Sr, \threshRd-\Sr\RPG\right),
\end{equation*}
and
\begin{multline*}
    \left\{
    \begin{alignedat}{10}
    &\threshRd - \Sr\RPG - \Sc \geq 0 \\
    &\frac{\Sc}{\threshC} - \Sr \geq \threshRd - \Sr\RPG - \Sc
    \end{alignedat}
    \right. \\
    \Leftrightarrow\quad \frac{\threshC\threshRd}{\threshC+1} - \Sr \frac{\threshC(\RPG-1)}{\threshC + 1} \leq \Sc \leq \threshRd - \RPG \Sr.
\end{multline*}
These conditions respectively lead to the blue and red areas on this plot (note that $\Srmin$, $\Srmax$ and $\Scmax$ are not illustrated):
\begin{center}
\adjustbox{width=0.9\linewidth}{
\begin{tikzpicture}
  \draw[->,thick] (0, 0) -- (6, 0) node[right] {$\Sr$};
  \draw[->,thick] (0,0) -- (0, 4) node[above] {$\Sc$};
    \draw[blue,thick,postaction={decoration={text along path,text={{\small $\threshC \Sr$}{}},text align={right},raise=0.2cm},decorate}] plot[domain=0:5.5] ({\x}, {0.4*\x});
    \draw[red,thick,postaction={decorate,decoration={text along path,text={{\small $\threshRd-\Sr \RPG$}{}},text align={left},raise=0.2cm}}] plot[domain=0:2.5] ({\x}, {3-1.2*\x});
    \draw[green,thick,postaction={decoration={text along path,text={{\small $\frac{\threshC\threshRd}{\threshC+1} - \Sr\:\frac{\threshC(\kappa-1)}{\threshC+1}$}{}},text align={right},raise=0.2cm},decorate}] plot[domain=0:5.5] ({\x}, {0.4/1.4*(3-0.2*\x});
    \fill[blue,opacity=0.1] plot[domain=0:1.875] (\x,3-1.2*\x) -- plot[domain=1.875:5.5] (\x,0.4*\x) -- (5.5,3.5) -- (0,3.5);
    \fill[red,opacity=0.1] plot[domain=0:1.875] (\x,0.4/1.4*(3-0.2*\x) -- plot[domain=1.875:0] (\x,3-1.2*\x);
    \draw[dashed] (1.875,0) -- (1.875,0.75);
    \draw[anchor=north] (1.875,0) node (n1) {{\normalsize $\frac{\threshRd}{\threshC+\RPG}$}};
    \draw[dashed] (0,0.75) -- (1.875,0.75);
    \draw[anchor=east] (0,0.75) node (n1) {{\normalsize $\frac{\threshC\threshRd }{\threshC+\RPG}$}};
\end{tikzpicture}}
\end{center}
and the areas $\setJRDCCPAPIC$ and $\setJRDCCPBPIC$ are then respectively rewritten as $\setJRDCCPA$ and $\setJRDCCPB$ given in \eqref{eq:JRDCCP_set}. Finally, the functions $\funcJRDCCPAPIC$ and $\funcJRDCCPBPIC$ of \eqref{eq:JRDCCP_func_IC} are replaced by $\funcJRDCCPA$ and $\funcJRDCCPB$ given in \eqref{eq:JRDCCP_func}, leading to \eqref{eq:JRDCCP_eq}.

\subsection{Detection probability}
\label{app:proof_DP}
\subsubsection{With interference cancellation}
The detection probability is obtained as
\begin{equation*}
\PDPIC(\gamma) = \lim_{\theta \rightarrow 0}\JRDCCPPIC(\theta;\gamma).
\end{equation*}
In that case, the area $\setJRDCCPBPIC$ is restricted to $\setPDPIC$ given in \eqref{eq:DP_set}, and the area $\setJRDCCPAPIC$ is restricted to $\setPDPIC^c$, where $\:\cdot\:^c$ denotes the complement set. In $\setPDPIC^c$, we have
\begin{equation*}
    \lim_{\theta \rightarrow 0} \Pr{\I|\Sr,\Sc}{\I \leq \frac{\Sc}{\threshC} - \resR} = 1.
\end{equation*}
In $\setPDPIC$, we have
\begin{align*}
   &\lim_{\theta \rightarrow 0} \Pr{\I|\Sr,\Sc}{\threshRd - \Sr\RPG - \resC \leq \I \leq \frac{\Sc}{\threshC} - \resR} \\
    &= \Pr{\I|\Sr,\Sc}{I \geq \threshRd - \Sr\RPG - \resC} \\
    &= \frac{1}{2} + \frac{1}{\pi}\int_0^\infty  \frac{1}{\tau}\:\Im{\LT_\I(-j\tau)\:\funcPDPIC(\tau)}\d\tau
\end{align*}
with $\funcPDPIC$ given in \eqref{eq:DP_func_IC}.

Finally, \eqref{eq:DP_eq_IC} is obtained knowing that
\begin{equation*}
\restesp{\setPDPIC^c}{1} = 1 - \restesp{\setPDPIC}{1}.
\end{equation*}

\subsubsection{Without interference cancellation}
Without interference cancellation, $\funcPDPIC$ and $\setPDPIC$ are rewritten as $\funcPD$ and $\setPD$ given in \eqref{eq:DP_func} and \eqref{eq:DP_set}.

\subsection{False alarm probability}
\label{app:proof_FAP}
\subsubsection{With interference cancellation}
The false alarm probability with interference cancellation is identical to the detection probability with interference cancellation, except that the radar echo is suppressed from the received power. In that case, $\funcPDPIC$ and $\setPDPIC$ are replaced by $\funcPFAPIC$ and $\setPFAPIC$ given in \eqref{eq:FAP_func_IC} and \eqref{eq:FAP_set_IC}.

\subsubsection{Without interference cancellation}
Without interference cancellation, $\funcPDPIC$ and $\setPFAPIC$ are rewritten as $\funcPFA$ and $\setPFA$ given in \eqref{eq:FAP_func} and \eqref{eq:FAP_set}.

\section{JRSCCP and derived metrics}
\subsection{JRSCCP metric}
\label{app:proof_JRSCCP}
\subsubsection{With interference cancellation}
The JRSCCP metric with interference cancellation can be developed as
\begin{align*}
    &\JRSCCP(\threshRs,\threshC) = \Pr{\Sr,\Sc,\I}{\frac{\Sr\RPG}{\Sc + \I} \geq \threshRs, \frac{\Sc}{\Sr + \I} \geq \threshC} \\
    &= \esp{\Sr,\Sc}{\Pr{\I|\Sr,\Sc}{\I \leq \frac{\Sr \RPG}{\threshRs}-\resC, \I \leq \frac{\Sc}{\threshC}-\resR}}.
\end{align*}
This probability is non zero either if
\begin{equation}
    \left\{
    \begin{alignedat}{10}
    &\frac{\Sr\RPG}{\threshRs} - \resC \geq 0\\
    &\frac{\Sc}{\threshC} - \resR \geq \frac{\Sr\RPG}{\threshRs} - \resC
    \end{alignedat}
    \right. , \label{app:condJRSCCPA}
\end{equation}
or
\begin{equation}
    \left\{
\begin{alignedat}{10}
    &\frac{\Sc}{\threshC} - \resR \geq 0 \\
    &\frac{\Sr\RPG}{\threshRs} - \resC \geq \frac{\Sc}{\threshC} - \resR
\end{alignedat}
    \right., \label{app:condJRSCCPB}
\end{equation}
leading to the areas $\setJRSCCPAPIC$ and $\setJRSCCPBPIC$ given in \eqref{eq:JRSCCP_set_IC} including the radar and communication power constraints.

In $\setJRSCCPAPIC$, applying the Gil-Pelaez theorem,
\begin{align*}
    &\Pr{\I|\Sr,\Sc}{\I\leq \frac{\Sr\RPG}{\threshRs} - \resC,\I\leq\frac{\Sc}{\threshC}-\resR} \\
    &= \Pr{\I|\Sr,\Sc}{\I \leq \frac{\Sr\RPG}{\threshRs} - \resC}\\
    &= \frac{1}{2} - \frac{1}{\pi} \int_0^\infty \frac{1}{\tau} \:  \Im{\LT_\I(-j\tau)\: \funcJRSCCPAPIC(\tau)} \d \tau
\end{align*}
with $\funcJRSCCPAPIC$ given in \eqref{eq:JRSCCP_func_IC}.

In $\setJRSCCPBPIC$, applying the Gil-Pelaez theorem,
\begin{align*}
    &\Pr{\I|\Sr,\Sc}{\I\leq \frac{\Sr\RPG}{\threshRs} - \resC,\I\leq\frac{\Sc}{\threshC}-\resR} \\
    &= \Pr{\I|\Sr,\Sc}{\I \leq \frac{\Sc}{\threshC} - \resR}\\
    &= \frac{1}{2} - \frac{1}{\pi} \int_0^\infty \frac{1}{\tau} \: \Im{\LT_\I(-j\tau)\: \funcJRSCCPBPIC(\tau)} \d \tau
\end{align*}
with $\funcJRSCCPBPIC$ given in \eqref{eq:JRSCCP_func_IC}, leading finally to \eqref{eq:JRSCCP_eq_IC}.

\subsubsection{Without interference cancellation}
Without interference cancellation, the JRSCCP is non zero only if
\begin{equation*}
    \left\{
    \begin{alignedat}{10}
    &\frac{\Sr \RPG}{\threshRs} - \Sc \geq 0 \\
    &\frac{\Sc}{\threshC} - \Sr \geq 0
    \end{alignedat}
    \right. \quad\Leftrightarrow\quad \threshC\Sr \leq \Sc \leq \frac{\Sr\RPG}{\threshRs},
\end{equation*}
which is fulfilled only if 
\begin{equation*}
    \threshC\threshRs \leq \kappa \quad\Leftrightarrow\quad [\threshC]_{\dB} + [\threshRs]_{\dB} \leq [\RPG]_{\dB}.
\end{equation*}
Then, \eqref{app:condJRSCCPA} and \eqref{app:condJRSCCPB} are respectively rewritten as
\begin{equation*}
    \left\{
    \begin{alignedat}{10}
    &\frac{\Sr\RPG}{\threshRs} - \Sc \geq 0\\
    &\frac{\Sc}{\threshC} - \Sr \geq \frac{\Sr\RPG}{\threshRs} - \Sc
    \end{alignedat}
    \right. \Leftrightarrow \Sr\frac{\threshC(\threshRs+\RPG)}{\threshRs(\threshC+1)} \leq \Sc \leq \frac{\Sr\RPG}{\threshRs},
\end{equation*}
and
\begin{equation*}
    \left\{
\begin{alignedat}{10}
    &\frac{\Sc}{\threshC} - \Sr \geq 0 \\
    &\frac{\Sr\RPG}{\threshRs} - \Sc \geq \frac{\Sc}{\threshC} - \Sr
\end{alignedat}
    \right. \Leftrightarrow \threshC\Sr \leq \Sc \leq \Sr\frac{\threshC(\threshRs+\RPG)}{\threshRs(\threshC+1)}.
\end{equation*}
These two cases respectively lead to the blue and red areas on this plot (note that $\Srmin$, $\Srmax$ and $\Scmax$ are not illustrated):
\begin{center}
\adjustbox{width=0.9\linewidth}{
\begin{tikzpicture}
  \draw[->,thick] (0, 0) -- (5, 0) node[right] {$\Sr$};
  \draw[->,thick] (0,0) -- (0, 4) node[above] {$\Sc$};
    \draw[blue,thick,postaction={decorate,decoration={text along path,text={{\small $\threshC \Sr$}{}},text align={right},raise=-0.3cm}}] plot[domain=0:4.5] (\x,0.2*\x);
    \draw[red,thick,postaction={decorate,decoration={text along path,text={{\small $\frac{\Sr\RPG}{\threshRs}$}{}},text align={right},raise=0.2cm}}] plot[domain=0:4.5] (\x,0.8*\x);
    \draw[green,thick,postaction={decorate,decoration={text along path,text={{\small $S_R\frac{\threshC(\threshRs+\RPG)}{\threshRs(\threshC+1)}$}{}},text align={right},raise=0.2cm}}] plot[domain=0:4.5] (\x,0.2/1.2*1.8*\x);
    \fill[blue,opacity=0.1] (0,0) --  plot[domain=0:4.5] (\x,0.2/1.2*1.8*\x) -- plot[domain=4.5:0] (\x,0.8*\x);
    \fill[red,opacity=0.1] (0,0) -- plot[domain=0:4.5] (\x,0.2*\x) -- plot[domain=4.5:0] (\x,0.2/1.2*1.8*\x);
\end{tikzpicture}
}
\end{center}
The areas $\setJRSCCPAPIC$ and $\setJRSCCPBPIC$ are then respectively rewritten as $\setJRSCCPA$ and $\setJRSCCPB$ given in \eqref{eq:JRSCCP_set}. Finally, the functions $\funcJRSCCPAPIC$ and $\funcJRSCCPBPIC$ of \eqref{eq:JRSCCP_func_IC} are respectively rewritten as $\funcJRSCCPA$ and $\funcJRSCCPB$ given in \eqref{eq:JRSCCP_func}, leading to \eqref{eq:JRSCCP_eq}.

\subsection{Coverage probability}
\label{app:proof_CP}
\subsubsection{With interference cancellation}
The coverage probability is obtained as 
\begin{equation*}
\PCPIC(\threshC) = \lim_{\threshRs \rightarrow 0} \JRSCCPPIC(\threshRs,\threshC).
\end{equation*}
In that case, $\setJRSCCPAPIC$ vanishes and $\setJRSCCPBPIC$ becomes $\setPCPIC$ given in \eqref{eq:CP_set_IC}, leading to \eqref{eq:CP_eq_IC}.

\subsubsection{Without interference cancellation}
Without interference cancellation, $\funcJRSCCPBPIC$ and $\setPCPIC$ are respectively rewritten as $\funcPC$ and $\setPC$ given in \eqref{eq:CP_func} and \eqref{eq:CP_set}.

\subsection{Success probability}
\label{app:proof_SP}
\subsubsection{With interference cancellation}
The success probability is obtained as 
\begin{equation*}
\PSPIC(\threshRs) = \lim_{\threshC \rightarrow 0} \JRSCCPPIC(\threshRs,\threshC).
\end{equation*}
In that case, $\setJRSCCPBPIC$ vanishes and $\setJRSCCPAPIC$ becomes $\setPCPIC$ given in \eqref{eq:CP_set_IC}.

\subsubsection{Without interference cancellation}
Without interference cancellation, $\funcJRSCCPAPIC$ and $\setPSPIC$ are respectively rewritten as $\funcPS$ and $\setPS$ given in \eqref{eq:SP_func} and \eqref{eq:SP_set}.

%%%%%%%%%%%%%%%%%%%%%%
\end{document}